\documentclass[11pt]{article}
\usepackage{graphicx}
\usepackage{amsmath}
\usepackage{amssymb}
\usepackage[british]{babel}
\usepackage{slashed} 
\usepackage{afterpage} 
\usepackage[hypertex]{hyperref}

\begin{document}

\begin{flushright}

\baselineskip=12pt

\hfill{ }\\

\end{flushright}

\begin{center}

{\Large\bf A numerical approach to
  harmonic non-commutative spectral field theory}

\vskip 1 cm

{\large Bernardino Spisso$^1$ and Raimar Wulkenhaar$^2$}
\\[3mm]
$^{1,2}$\,\emph{Mathematisches Institut der Westf\"alischen
  Wilhelms-Universit\"at\\
Einsteinstra\ss{}e 62, D-48149 M\"unster, Germany}

\footnotetext[1]{nispisso@tin.it}
\footnotetext[2]{raimar@math.uni-muenster.de}

\vspace{1cm}

\end{center}
\vskip 1 cm
\begin{abstract}
  We present a first numerical investigation of a non-commutative
  gauge theory defined via the spectral action for Moyal space with
  harmonic propagation. This action is approximated by finite
  matrices.  Using Monte Carlo simulation we study various quantities
  such as the energy density, the specific heat density and some order
  parameters, varying the matrix size and the independent parameters
  of the model. We find a peak structure in the specific heat which
  might indicate possible phase transitions. However, there are mathematical
  arguments which show that the limit of infinite matrices can be quite
  different from the original spectral model.  
\end{abstract}

\section{Introduction }

Quantum field theory on noncommutative spaces
\cite{Douglas:1999ge,Szabo:2001kg,Wulkenhaar:2006si} is an active
subject of research. The most-studied noncommutative spaces are the
Moyal space \cite{GraciaBondia:1987kw} and fuzzy spaces
\cite{Madore:1992bw}. Fuzzy spaces are matrix approximations of
manifolds and as such ideal for numerical investigations similar to
non-perturbative quantum field theory on the lattice. In this paper we
focus on the Moyal space, which is a continuous deformation of
$\mathbb{R}^d$ for which the usual Fourier techniques of perturbative
quantum field theory are available.  It turned out that a
renormalisable quantum field theory on $\mathbb{R}^d$ is, in most
cases, no longer renormalisable on $d$-dimensional Moyal space due to
a phenomenon called ultraviolet/infrared mixing
\cite{Minwalla:1999px}. In \cite{Grosse:2004yu} it was discovered that
for the $\varphi^4$-model on 4-dimensional Moyal space the
UV/IR-mixing generates an additional marginal coupling which
corresponds to a harmonic oscillator potential for the
free scalar field. The resulting action
\begin{equation}
S[\varphi]=\int d^4x \left(\frac{1}{2} \varphi \star 
(-\Delta+\Omega^2 \tilde{x}^2 + \mu^2)\star\varphi 
+ \frac{\lambda}{4}\varphi \star \varphi \star \varphi 
\star \varphi\right)(x) 
\label{action}
\end{equation}
was then shown to be perturbatively renormalisable to all orders in
$\lambda$. In (\ref{action}), $\tilde{x} = 2\Theta^{-1} \cdot x$,
where $\Theta$ is the deformation matrix defining the Moyal product.
See also
\cite{Grosse:2005da,Rivasseau:2005bh,Gurau:2005gd,Gurau:2006yc,Rivasseau:2007ab}.
Moreover, the frequency parameter can be restricted to $\Omega \in
[0,1]$ by Langmann-Szabo duality \cite{Langmann:2002cc}.

The renormalisability of the action (\ref{action}) raises the question
whether a harmonic term can also render Yang-Mills theory
renormalisable on Moyal space (recall that the usual Yang-Mills action
on Moyal space has the same UV/IR-mixing problem
\cite{Matusis:2000jf}). Yang-Mills theories in noncommutative geometry
\cite{Connes} are naturally obtained from the spectral action
principle \cite{Chamseddine:1996zu} relative to an appropriate Dirac
operator. In \cite{Gayral:2003dm} it was shown that Moyal space (with
usual Dirac operator) is a (non-compact) spectral triple; its
corresponding spectral action was computed in \cite{Gayral:2004ww}.
To obtain a gauge theory with harmonic oscillator potential via the
spectral action principle, a differential square root of the harmonic
oscillator Hamiltonian is necessary as Dirac operator. In absence of
such a Dirac operator, in \cite{de Goursac:2007gq,Grosse:2007dm} an
effective gauge model was constructed as the one-loop effective action
of complex harmonic noncommutative quantum $\varphi^4$-theory in a
classical external gauge field. As a result, the noncommutative
Yang-Mills Lagrangian is extended by two terms $X_\mu\star X^\mu$ and
$(X_\mu\star X^\mu)^2$, where $X_\mu=\frac{1}{2} \tilde{x}_\mu+A_\mu$
is the `covariant coordinate'.

A first outline of a candidate spectral triple for harmonic oscillator
Moyal space was given in \cite{Grosse:2007jy}. Additionally, in
\cite{Grosse:2007jy} the linear and quadratic terms of the spectral
action for a $U(1)$-Yang-Mills-Higgs model were computed and then
extended by gauge invariance. Thereby the appearance of $X_\mu\star
X^\mu$ was traced back to a deep entanglement of gauge and Higgs
fields in a unified potential $(\alpha X_\mu \star X^\mu + \beta
\bar{\varphi} \star \varphi - 1)^2$, with $\alpha, \beta \in
\mathbb{R}^+$.

It turned out that the candidate spectral triple proposed in
\cite{Grosse:2007jy} was the shadow of a new class of non-compact
spectral triples with finite volume \cite{Wulkenhaar:2009pv}. The
spectral geometry of Moyal space with harmonic propagation, which
falls into this class, was fully worked out in \cite{Gayral:2011vu}.
There are in fact two (even, real) spectral triples
$(\mathcal{A}_\star,\mathcal{H},\mathcal{D}_\bullet,\Gamma,J)$, with
$\bullet\in \{1,2\}$, for the $d$-dimensional Moyal algebra
$\mathcal{A}_\star$ and differential square roots $\mathcal{D}_\bullet$ of
the harmonic oscillator Hamiltonian. The spectral triples are of
metric dimension $d$ and KO-dimension $2d$, have simple dimension
spectrum consisting of the integers $\leq d$, and satisfy all regularity and
compatibility requirements of spectral triples.
Additionally, the spectral action was rigorously computed in
\cite{Gayral:2011vu}, i.e.\ with H\"older type estimates for the
remainder of the asymptotic expansion and with inclusion of the real
structure $J$.

A completely new feature of the spectral action
\cite{Gayral:2011vu,Grosse:2007jy} (and also of the effective action
\cite{de Goursac:2007gq,Grosse:2007dm}) is that the expansion of
$X_\mu \star X^\mu$ and its square produces a term which is
\emph{linear in the gauge field $A$}. This means that the vacuum,
i.e.\ the solution of the classical field equations, is no longer
taken at $A_\mu=0$ (or more generally at a flat connection
$F_{\mu\nu}=0$) but at some non-constant value for the gauge field. A
first discussion of the vacuum structure of this type of gauge models
was given in \cite{deGoursac:2008rb}. It turned out that generically there are
infinitely many vacuum solutions. Some of them were exposed, but it
was not possible to give reasonable argument for the right solution.
In particular, it became completely impossible to study the gauge
model as a perturbative quantum field theory.

This is the point where the numerical treatment comes into play. The
standard method of numerical quantum field theory is to approximate
the space by discrete points, for example using a lattice
approximation and then calculate the observables over that set of
points \cite{lattice}.  For Moyal space a position space approximation
is not suitable due to the oscillator factor of the Moyal product.
Instead, we shall use the matrix Moyal base (which was already used in
the first renormalisation proof \cite{Grosse:2004yu} of
$\varphi^4$-model), restrict it to finite matrices and perform a Monte
Carlo simulation of the resulting action.  In this way we will study
some statistical quantities such as energy density and specific heat,
varying the parameters $\Omega, \frac{\chi_{-1}}{\chi_0}, \alpha$ of
the model and gathering some information on the various contributions
of the fields to the action.  The simulations are quite cumbersome due
the complexity of the action and the number of independent matrices to
handle. Nevertheless we are able to get an acceptable balance between
the computation precision and the computation time. For the
simulations we apply a standard Metropolis-Monte Carlo algorithm
\cite{Metro} with various estimators for the error and for the
autocorrelation time of the samples.  The range of parameters is
chosen to avoid problems with the thermalisation process, thus
permitting a relative small number of Monte Carlo steps to compute
independent results from the initial conditions.

We are eventually interested in the continuum limit which corresponds
to matrices of infinite size. We thus compute our observables such as the
energy density for various matrix sizes and then look for a stabilization of 
these observables as the matrix size increases. 
The specific heat, which is a measure of the dispersion of the energy,
will be used to identify possible phase transitions in form of 
peaks of the specific heat at increasing matrix size.

\section{Four-dimensional harmonic Yang-Mills model}

The harmonic Yang-Mills model is defined as the spectral action
resulting from the spectral triples
$(\mathcal{A}_\star,\mathcal{H},\mathcal{D}_\bullet,\Gamma,J)$, with
$\bullet\in \{1,2\}$, analysed in \cite{Gayral:2011vu}. The 
Moyal algebra $\mathcal{A}_\star$ is the space of
Schwartz class functions on $\mathbb{R}^4$ equipped with the product 
\begin{align}
f\star g(x) = \int_{\mathbb{R}^4\times \mathbb{R}^4} \frac{dy
 \,dk}{(2\pi)^4} \;
f(x{+}\tfrac{1}{2}
\Theta \cdot k) \,g(x{+}y)\, \mathrm{e}^{i\langle k,y\rangle} \;.
\label{Moyal}
\end{align}
Where $\langle k,y\rangle$ is the scalar product defined as $k_\mu y^\mu$. The unbounded selfadjoint operators $\mathcal{D}_\bullet$ on the
Hilbert space $\mathcal{H}$ are differential square roots of the
harmonic oscillator Hamiltonian $H=-\partial_\mu\partial^\mu +
\tilde{\Omega}^2 x_\mu x^\mu$ of frequency $\tilde{\Omega}$, i.e.\
$\mathcal{D}_\bullet^2=H-(-1)^\bullet \tilde{\Omega}\Sigma$, for a
certain spin matrix $\Sigma$. If $L_\star(f)$ denotes left Moyal
multiplication with a function $f\in \mathcal{A}_\star$, then one has 
\begin{align}
[\mathcal{D}_1, L_\star (f)]&=
L_\star(i \partial_\mu f)\otimes \Gamma^\mu\;,\qquad &
[\mathcal{D}_2, L_\star (f) ] &=
L_\star(i \partial_\mu f)\otimes \Gamma^{\mu+4}\;, 
\label{Gamma}
\end{align}
where the matrices $\Gamma_1,\dots,\Gamma_8$ satisfy the
anticommutation relations
\begin{align}
  \{ \Gamma^\mu,\Gamma^\nu\} &=\{ \Gamma^{\mu+4},\Gamma^{\nu+4}\} = 2
  (g^{-1})^{\mu\nu}\;,\quad \{ \Gamma^\mu,\Gamma^{\nu+4}\}=0\;,
\end{align}
relative to an induced metric
$g=(\mathrm{id}-\frac{1}{4}\tilde{\Omega}^2\Theta^2)^{-1}$. 
The grading is $\Gamma=\Gamma_1\cdots\Gamma_8$, and the real structure
satisfy $J\mathcal{D}_\bullet J^{-1}=\mathcal{D}_\bullet$ and 
$JL_\star(f) J^{-1}=R_\star(\bar{f})$, where $R_\star$ denotes right
Moyal multiplication. 

In order to implement the Higgs mechanism \`a la Connes-Lott
\cite{Connes:1990qp} one considers the product of the spectral triple
$(\mathcal{A}_\star,\mathcal{H},\mathcal{D}_1,\Gamma,J)$ with the
finite Higgs spectral triple $(\mathbb{C} \oplus
\mathbb{C},\mathbb{C}^2,M \sigma_1, J_f)$, where $\sigma_1$ is a Pauli
matrix, $J_f$ any matricial real structure and $M > 0$.  Then, a
self-adjoint fluctuation $A=\sum a_i[\mathcal{D},b_i]$ of the total
Dirac operator $\mathcal{D}=(\mathcal{D}_1 \otimes 1 + \Gamma \otimes
M\sigma_1)$ to give $\mathcal{D}_A=\mathcal{D}+A+{\rm J}A{\rm
  J}^{-1}$, for ${\rm J}=J\otimes J_f$, is of the form
\begin{equation}
A =\left(\begin{array}{cc} 
\Gamma^\mu L_\star(A_\mu )& \Gamma L_\star(\phi) \\
\Gamma L_\star (\bar{\phi}) & \Gamma^\mu L_\star (B_\mu ) \end{array}\right)\;,
\label{A-fluct}
\end{equation}
for the components $A_\mu,B_\mu\in \mathcal{A}_\star$ of two real one-forms and a
complex scalar $\phi \in \mathcal{A}_\star$. Using $\mathcal{D}_2$ 
instead of $\mathcal{D}_1$ amounts to replace $\Gamma^\mu$ 
by $\Gamma^{\mu+4}$.

The spectral action principle \cite{Chamseddine:1996zu} asserts that
the bosonic action of a field theory with fermionic Dirac operator
$\mathcal{D}_A$ has the form
\begin{align}
S(\mathcal{D}_A) = \mathrm{Tr}(\chi(\mathcal{D}_A^2))\;,
\end{align}
where $\chi$ is a smooth approximation of the characteristic function on
$[0,\Lambda^2]$, for some scale parameter $\Lambda$. For the
fluctuation (\ref{A-fluct}), the part of the spectral action which is
relevant and marginal for $\Lambda \to \infty$ has been explicitly
computed in \cite{Gayral:2011vu}, for general effective metric $g$.
This computation involved Laplace transformation, Duhamel expansion
with H\"older-type estimates for the remainder and explicit use of the
Mehler kernel for the harmonic oscillator Hamiltonian. For a special
choice of the noncommutativity matrix $\Theta^2=-\theta^2 \mathrm{id}$,
the result of \cite{Gayral:2011vu} takes in terms of $\Omega:=
\frac{\theta \tilde{\Omega}}{2}$, the moments $\Lambda^{2n} \chi_{-n}:
=\int_0^\infty ds \;s^{n-1}\chi(s)$ of the ``characteristic function''
and $\chi_0:=\chi(0)$ the form
\begin{align}
&S_\Lambda(\mathcal{D}_A)
=\frac{\theta^4\Lambda^8}{8 \Omega^{4}} \chi_{-4}
-\frac{ M^2 \theta^4 \Lambda^6}{8 \Omega^{4}} \chi_{-3}
+ \Big(\frac{M^4\theta^4 \Lambda^4}{16 \Omega^4}
+\frac{8\theta^2 \Lambda^4}{12 \Omega^2}\Big) \chi_{-2}
- \Big(\frac{M^6\theta^4 \Lambda^2}{48 \Omega^4}
+\frac{2 M^2\theta^2 \Lambda^2}{3 \Omega^2} \Big)  \chi_{-1}
\nonumber
\\
&\qquad \qquad + \Big(\frac{52}{45} +\frac{M^8\theta^4}{192 \Omega^4}
+\frac{M^4\theta^2}{3\Omega^2} \Big) \chi_0
\nonumber
\\
&+\frac{\chi_0}{\pi^2(1+\Omega^2)^2} \int d^4x 
\bigg\{
2 (1+\Omega^2) D_\mu \phi \star \overline{D^\mu\phi}
\nonumber
\\
&+ \Big(\phi{\star} \bar{\phi} +M(\phi {+} \bar{\phi})
+ \frac{4 \Omega^2}{1{+}\Omega^2} \tilde{X}_{A\mu}{\star} \tilde{X}^\mu_A
+M^2 - \frac{\chi_{-1}}{\chi_0}\Lambda^2 \Big)^2
- \Big(
\frac{4 \Omega^2}{1{+}\Omega^2} \tilde{X}_{0\mu}{\star} \tilde{X}^\mu_0
+M^2 - \frac{\chi_{-1}}{\chi_0}\Lambda^2 \Big)^2
\nonumber
\\
&+\Big(\bar{\phi} {\star} \phi +M(\phi {+} \bar{\phi})
+ \frac{4 \Omega^2}{1{+}\Omega^2} \tilde{X}_{B\mu}{\star} \tilde{X}^\mu_B
+M^2 - \frac{\chi_{-1}}{\chi_0}\Lambda^2 \Big)^2
- \Big(
\frac{4 \Omega^2}{1{+}\Omega^2} \tilde{X}_{0\mu}{\star} \tilde{X}_0^\mu
+M^2 - \frac{\chi_{-1}}{\chi_0}\Lambda^2 \Big)^2
\nonumber
\\
&
+\Big( \frac{(1+\Omega^2)^2}{2} 
- \frac{(1-\Omega^2)^4}{6 (1+\Omega^2)^2} \Big)
\big(F^A_{\mu\nu} \star F^{A\mu\nu}
+F^B_{\mu\nu} \star F^{B\mu\nu}\big)\bigg\}
\nonumber
\\
& + \mathcal{O}(\Lambda^{-1})\;.
\label{SpecAct}
\end{align}
Here, $D_\mu\phi=\partial_\mu \phi -i A_\mu\star \phi+i \phi \star
B_\mu -i M(A_\mu-B_\mu)$ is the covariant derivative of the scalar
field, $F^A_{\mu\nu}: = \partial_\mu A_\nu - \partial_\nu A_\mu
-i(A_\mu \star A_\nu - A_\nu \star A_\mu)$ the field strength of $A$
and similarly $F^B_{\mu\nu}$ the field strength of $B$. Moreover,
$\tilde{X}_{A\mu}:=\tilde{X}_{0\mu}+A_\mu$ and
$\tilde{X}_{B\mu}:=\tilde{X}_{0\mu}+B_\mu$ are the covariant
derivatives of $A$ and $B$, respectively, where
$\tilde{X}_{0\mu}:=\frac{\tilde{x}_\mu}{2} =(\Theta^{-1})_{\mu\nu}
x^\nu$. The remarkable outcome of the spectral action (\ref{SpecAct})
is that the Higgs field $\phi$ \emph{and} the gauge fields $A,B$
appear together in a unified potential. In this way, also the gauge
field shows a non-trivial vacuum structure. Besides, the action is
invariant under $U(\mathcal{A}_\star) \times U(\mathcal{A}_\star)$
transformations:
\begin{equation}
(\phi+M) \mapsto u_A\star(\phi+M)\star\overline{u_B}, \quad 
\tilde{X}^A_{\mu} \mapsto u_A\star \tilde{X}^\mu_A\star
\overline{u_A}, \quad 
\tilde{X}^\mu_B\mapsto u_B\star\tilde{X}^\mu_B\star \overline{u_B} \;.
\label{gauge}
\end{equation}

\section{Discretisation by Moyal base}

The 2-dimensional Moyal algebra with deformation parameter $\theta>0$
has a natural basis of eigenfunctions $f_{mn}$ of the harmonic
oscillator, where $m, n \in \mathbb{N}$.  These are given in radial
coordinates by
\begin{equation}
f_{mn}(\rho \cos \varphi,\rho \sin \varphi) 
= 2(-1)^m \sqrt{\frac{m!}{n!}} e^{i\varphi(n-m)} 
\left(\sqrt{\frac{2}{\theta}}\rho\right)^{n-m} 
e^{-\frac{\rho^2}{\theta}}L_{m}^{n-m}\left(\frac{2}{\theta}\rho^2\right) 
\label{mb}
\end{equation}
and satisfy 
\begin{align}
(f_{mn} \star f_{kl} )(x) & = \delta_{nk} f_{ml} (x) \label{star-rule}
\\
\int d^2 x f_{mn}(x)&= 2\pi\theta\delta_{mn}\;,
\end{align}
see \cite{GraciaBondia:1987kw,Gayral:2003dm} for details. The
expansion of Schwartz functions on $\mathbb{R}^4$ in the Moyal base,
\begin{equation}
\mathcal{A}_\star \ni 
a = a(x_0,\dots,x_3) = \sum_{m_1,m_2,n_1,n_2\in \mathbb{N}}
a_{\genfrac{}{}{0pt}{}{m_1n_1}{m_2n_2}}
f_{m_1n_1}(x_0,x_1)f_{m_2n_2}(x_2,x_3)\;,
\end{equation}
then provides an isomorphism of Fr\'echet spaces between
$\mathcal{A}_\star $ and the space of rapidly decreasing double sequences 
$(a_{mn})_{m,n\in \mathbb{N}^2}$ equipped with the 
family of seminorms 
\begin{align}
p_k((a_{mn})_{m,n\in \mathbb{N}^2})
:= \sum^\infty_{m,n \in \mathbb{N}^2}
\left((2|m|+1)^{2k} (2|n|+1)^{2k} |a_{mn}|^2
\right)^{\frac{1}{2}}\;,\qquad
|m|:=m_1+m_2\;.
\label{Frechet}
\end{align}
According to (\ref{star-rule}), Moyal product and integral reduce in
the $(f_{mn})$-basis to product and trace of infinite
$\mathbb{N}^2$-labeled matrices, with convergent index sums due to
(\ref{Frechet}).  By duality, the covariant derivatives $X^A_{\mu}$
and $X^B_\mu$ can also be expanded in the $(f_{mn})$-basis, but the
expansion coefficients $X^A_{\mu\genfrac{}{}{0pt}{}{m_1n_1}{m_2n_2}}$,
$X^B_{\mu\genfrac{}{}{0pt}{}{m_1n_1}{m_2n_2}}$ diverge for $m_i,n_i\to
\infty$.

To any $a \in \mathcal{A}_\star$ we can associate a sequence
$(a^N)_{N\in \mathbb{N}}$ of cut-off matrices 
\[
a^N_{\genfrac{}{}{0pt}{}{m_1n_1}{m_2n_2}}=\left\{
\begin{array}{cl}
a_{\genfrac{}{}{0pt}{}{m_1n_1}{m_2n_2}} &\qquad\text{if
}\max(m_1,m_2,n_1,n_2)\leq N \;,
\\
0 & \qquad\text{else}\;.
\end{array}\right.
\]
Then, $(a^N)$ is a Cauchy sequence in any of the semi-norms $p_k$ and
converges to $a$ in the Fr\'echet topology of $\mathcal{A}_\star$. 

In quantum field theory we are confronted with the converse problem.
To deal with divergences, a regularisation has to be introduced which
restricts the system to a finite number of degrees of freedom. After
re-normalisation from bare to physical quantities one has to show that
the limit to an infinite number of degrees of freedom is well-defined.
In our case, the natural regularisation is to restrict the matrix
indices to $m_i\leq N$, which corresponds to a cut-off in the energy.
Even if we could solve the renormalisation problem, the removal of the
cut-off, i.e.\ the limit $N\to \infty$ to infinite matrices, will
fail: A sequence of $(N\times N)$-matrix algebras does not converge in
the Fr\'echet topology.

Fortunately, in quantum field theory we are interested in the
convergence of correlation functions, and not of matrix algebras.  The
path integral in usual quantum field theories is over random walks and
not over smooth field configurations. It seems not impossible
(although we cannot prove it) that quantum correlation functions are
less sensitive to the topology of the underlying classical field
theory. In our case, we make the (not verifiable) hypothesis that the
cut-off correlation functions to be computed carry some information about
the original smooth model.



Using the identities $D_\mu\phi =i(\phi+M)\star\tilde{X}_{B\mu}
-i \tilde{X}_{A\mu}\star(\phi+M)$ and
$F^A_{\mu\nu}=-i [X^A_\mu,X^A_{\nu}]+i[X^0_\mu,X^0_{\nu}]$ (and similarly
for $F^B_{\mu\nu}$), and ignoring all contributions of $X^0_\mu$ which
for finite matrices yields some finite number,
we can recast the restriction of the action \eqref{SpecAct} to finite
matrices in the following form:
\begin{align}
S(\phi,\tilde{X}_A,\tilde{X}_B) 
&= \frac{1}{(1+\Omega^2)^2} \mathrm{Tr}\Bigg\{
\left(\frac{\left(1-\Omega^2\right)^2}{2}
-\frac{\left(1+\Omega^2\right)^4}{6\left(1+\Omega^2\right)^2}\right)
\Big(\left[\tilde{X}_{A\mu},\tilde{X}_{A\nu}\right]_\star
\left[\tilde{X}_{A}^\mu,\tilde{X}_{A}^\nu\right]_\star  \nonumber 
\\
&+\left[\tilde{X}_{B\mu},\tilde{X}_{B\nu}\right]_\star
\left[\tilde{X}_{B}^\mu,\tilde{X}_{B}^\nu\right]_\star\Big)\nonumber 
\\
&+ \left(\phi\star\bar{\phi}+M(\phi {+} \bar{\phi})
+\frac{4\Omega^2}{1+\Omega^2}\tilde{X}_A^\mu\star\tilde{X}_{A\mu} +M^2
-\Lambda^2\frac{\chi_{-1}}{\chi_0}\right)^2 \nonumber 
\\
&+\left(\bar{\phi}\star\phi +M(\phi {+} \bar{\phi})
+\frac{4\Omega^2}{1+\Omega^2}\tilde{X}_B^\mu\star\tilde{X}_{B\mu} +M^2
-\Lambda^2\frac{\chi_{-1}}{\chi_0} \right)^2  \nonumber 
\\
&+2(1+\Omega^2)\left((\phi+M)\star\tilde{X}_{B\mu}
-\tilde{X}_{A\mu}\star(\phi+M)\right)\nonumber 
\\
& \left((\bar{\phi}+M)\star\tilde{X}_{A}^\mu
-\tilde{X}_{B}^\mu\star(\bar{\phi}+M)\right)\Bigg\}\;.
\label{S0}
\end{align}

The restriction to finite matrices shows crucial differences to the
smooth model. Only these differences make the numerical simulations
possible with the drawback of the serious possiblility that our results  can deviate from the original
smooth model.
\begin{enumerate}
\item The action (\ref{S0}) has an obvious family of minima given
appropriate multiples of the identity matrices. We thus define
\begin{align}
\phi +M &= \psi +\Lambda 
\sqrt{\frac{\chi_{-1}}{\chi_0}}\cos\alpha\textbf{I} \ \\
\tilde{X}_{A\mu}&= Y_{A\mu} +\frac{1}{2}\Lambda 
\sqrt{\frac{\chi_{-1}}{\chi_0}}\sqrt{\frac{2\Omega^2}{(1+\Omega^2)}}
\textbf{I}_\mu\sin\alpha  \\
\tilde{X}_{B\mu}&= Y_{B\mu} +\frac{1}{2}\Lambda 
\sqrt{\frac{\chi_{-1}}{\chi_0}}
\sqrt{\frac{2\Omega^2}{(1+\Omega^2)}}\textbf{I}_\mu\sin\alpha  \;.
\end{align}
Note that the corresponding minimum configurations for $A_\mu,B_\mu$ 
explicitly violate, in the limit $N\to\infty$, the Fr\'echet condition. 

Substituting the previous fields into \eqref{S0} we get a positive
action with minimum in zero:
\begin{eqnarray}
&S(\psi,Y_A,Y_B)&=  \frac{1}{(1+\Omega^2)^2}\mathrm{Tr}\Bigg\{D\Big(\left[Y_{A\mu},Y_{A\nu}\right]\left[Y_{A}^\mu,Y_{A}^\nu\right] 
+ \left[Y_{B\mu},Y_{B\nu}\right]\left[Y_{B}^\mu,Y_{B}^\nu\right]\Big) \nonumber \\
& &+ \left(\psi\bar{\psi}+\mu\cos\alpha(\psi+\bar{\psi})+ CY_A^\mu Y_{A\mu} +\mu \textbf{I}^\mu Y_{A\mu}\sin\alpha \right)^2  \nonumber \\
& &+ \left(\bar{\psi}\psi +\mu\cos\alpha(\psi+\bar{\psi})+ CY_B^\mu Y_{B\mu} +\mu \textbf{I}^\mu Y_{B\mu}\sin\alpha  \right)^2 \nonumber \\ 
& &+2(1+\Omega^2)\left((Y_{B\mu}-Y_{A\mu})\mu\cos\alpha +\psi Y_{B\mu}-Y_{A\mu}\psi\right)\nonumber \\
& &\left((Y_{A}^\mu-Y_{B}^\mu)\mu\cos\alpha+\bar{\psi} Y_{A}^\mu -Y_{B}^\mu\bar{\psi}\right)\Bigg\},
\label{Sf} \end{eqnarray}
with 
\begin{align}
C&=\frac{1+\Omega^2}{4\Omega^2}\;,& 
D&=\frac{\left(1-\Omega^2\right)^2}{2}
-\frac{\left(1+\Omega^2\right)^4}{6\left(1+\Omega^2\right)^2} \;, & 
\Lambda^2 \frac{\chi_{-1}}{\chi_0}&=\mu^2\;.
\end{align}

\item For finite matrices, the $\mathbb{N}^2$-indexed double sequences
  can be written as tensor products of ordinary matrices, 
\begin{equation}
X_{\genfrac{}{}{0pt}{}{m_1n_1}{m_2n_2}}= \sum_{i=1}^K 
X^i_{m_1n_1} \otimes X^i_{m_2n_2} \;.
\label{tensorproduct}
\end{equation}
Since the matrix product and trace also
factor into these independent components, the action factors into 
$S=\sum_{i=1}^K
S(\psi^{1i},Y_A^{1i},Y_B^{1i})S(\psi^{2i},Y_A^{2i},Y_B^{2i})$. 
Then, regarding all 
$\psi^{1i},Y_A^{1i}$, $Y_B^{1i},\psi^{2i},Y_A^{2i},Y_B^{2i}$
as random variables over which to integrate in the partition function,
the partition function factors, too:
\begin{align}
&\int
\mathcal{D}(\psi^{11},Y_A^{11},Y_B^{11},\psi^{21},Y_A^{21},Y_B^{21})
\cdots 
\mathcal{D}(\psi^{1K},Y_A^{1K},Y_B^{1K},\psi^{2K},Y_A^{2K},Y_B^{2K})
\;e^{-S}
\nonumber
\\
&=\bigg(
\int \mathcal{D}(\psi^{1i},Y_A^{1i},Y_B^{1i},\psi^{2i},Y_A^{2i},Y_B^{2i})
\;e^{-S(\psi^{1i},Y_A^{1i},Y_B^{1i})\cdot S(\psi^{2i},Y_A^{2i},Y_B^{2i})}
\bigg)^K\;.
\end{align}
We may therefore restrict ourselves to $K=1$. Now the discretized action is invariant under the same transformations
\eqref{gauge}, but now with $(u_A,u_B) \in U(\mathbb{M}_N) \times U(\mathbb{M}_N)$.
For the limit $N\to\infty$ we would have $K\to \infty$ and therefore problems with  convergence.

\item Instead of integrating in the partition function over all
  gauge-equivalence classes of $\psi,Y_A,Y_B$ as required, we follow
  the usual matrix model philosophy\footnote{In scalar 1-matrix models
    the gauge-equivalence classes are the configurations of
    eigenvalues. Nevertheless it is custom to integrate over
    \emph{all} matrices. The integration over the gauge group produces
    a measure for the eigenvalues which is given by the square of the
    Vandermonde determinant. In 2-matrix models the gauge equivalence
    classes are the eivenvalues together with a unitary matrix which
    describes the relative orientation of the eigenbases. The full
    matrix integration can be reduced to an integration over the
    eingenvalues thanks to the formula of Itzykson-Zuber and
    Harish-Chandra.} and integrate over \emph{all} matrices
  $\psi,Y_A,Y_B$. A reduction of this 10-matrix model to
    gauge-equivalence classes seems rather hopeless.

\end{enumerate}

It is convenient to pass, for each factor in the tensor product
(\ref{tensorproduct}), to complex matrices \cite{deGoursac:2008rb}:
\begin{align}
Z_0 & ={Y}^A_0+i{Y}^A_1, & \bar{Z}_0
&={Y}^A_0-i{Y}^A_1  
\nonumber \\
 Z_1&={Y}^B_0+i{Y}^B_1, & \bar{Z}_1 &={Y}^B_0-i{Y}^B_1 \nonumber \\
 Z_2&={Y}^A_2+i{Y}^A_2, & \bar{Z}_2 &={Y}^A_2-i{Y}^A_3  \nonumber \\
 Z_3&={Y}^B_2+i{Y}^B_3, & \bar{Z}_3 &={Y}^B_2-i{Y}^B_3  
\label{Z-sub}
\end{align}
The convention that the bar denotes the hermitian conjugate will also
be used for the complex matrix $\psi$.
In the end  using the substitutions \eqref{Z-sub} and after some simple manipulations, the discretised action is: 
\begin{equation}
S_4=\frac{1}{(1+\Omega^2)}\operatorname{Tr}\left(\mathcal{L}_F+\mathcal{L}_{V_0}+\mathcal{L}_{V_1}+\mathcal{L}_{D_0}\bar{\mathcal{L}}_{D_0}
  +\mathcal{L}_{D_1}\bar{\mathcal{L}}_{D_1}+\mathcal{L}_{D_2}\bar{\mathcal{L}}_{D_2}+\mathcal{L}_{D_3}\bar{\mathcal{L}}_{D_3}\right) \;,
\label{S4}
\end{equation}
with 
{\allowdisplaybreaks[4]
\begin{align}
\mathcal{L}_{F}&=\frac{D}{2}\Big(\left[\bar{Z}_0,Z_0\right]^2 +\left[\bar{Z}_1,Z_1\right]^2 + 
\frac{1}{4}\Big(\left[Z_0+\bar{Z}_0,Z_2-\bar{Z}_2\right]^2-\left[Z_0+\bar{Z}_0,Z_2+\bar{Z}_2\right]^2\nonumber 
\\*
&+ \left[Z_0-\bar{Z}_0,Z_2+\bar{Z}_2\right]^2-
\left[Z_0-\bar{Z}_0,Z_2-\bar{Z}_2\right]^2
-\left[Z_1+\bar{Z}_1,Z_3+\bar{Z}_3\right]^2\nonumber 
\\*
&+ \left[Z_1+\bar{Z}_1,Z_3-\bar{Z}_3\right]^2
+\left[Z_1-\bar{Z}_1,Z_3+\bar{Z}_3\right]^2
-\left[Z_1-\bar{Z}_1,Z_3-\bar{Z}_3\right]^2\Big)\Big)\nonumber 
\\
\mathcal{L}_{V_0}&=\big(\psi\bar{\psi}+\mu\cos\alpha(\psi+\bar{\psi})+
\frac{1}{2}\left(\left\{\bar{Z}_0,Z_0\right\}
  +\left\{\bar{Z}_2,Z_2\right\}\right)\nonumber 
\\*
&+\frac{\mu\sin\alpha}{2\sqrt{C}}((-1+i)(Z_0+Z_2)+(1+i)(\bar{Z}_0+\bar{Z}_2))\big)^2
\nonumber 
\\
\mathcal{L}_{V_1}&=\big(\bar{\psi}\psi+\mu\cos\alpha(\psi+\bar{\psi})+
\frac{1}{2}\left(\left\{\bar{Z}_1,Z_1\right\}
  +\left\{\bar{Z}_3,Z_3\right\}\right)\nonumber 
\\*
&+
\frac{\mu\sin\alpha}{2\sqrt{C}}((-1+i)(Z_1+Z_3)
+(1+i)(\bar{Z}_1+\bar{Z}_3))\big)^2
\nonumber 
\\
\mathcal{L}_{D_0}&=
\sqrt{2(1+\Omega^2)}\left(\mu\cos\alpha(Z_1+\bar{Z}_1-Z_0-\bar{Z}_0 )
  + \psi(Z_1+\bar{Z}_1)-(Z_0+\bar{Z}_0)\psi\right)\nonumber 
\\
\mathcal{L}_{D_1}&=
\sqrt{2(1+\Omega^2)}\left(\mu\cos\alpha(Z_1-\bar{Z}_1-Z_0+\bar{Z}_0 )
  + \psi(Z_1-\bar{Z}_1)-(Z_0-\bar{Z}_0)\psi\right)\nonumber 
\\
\mathcal{L}_{D_2}&=
\sqrt{2(1+\Omega^2)}\left(\mu\cos\alpha(Z_3+\bar{Z}_3-Z_2-\bar{Z}_2 )
  + \psi(Z_3+\bar{Z}_3)-(Z_2+\bar{Z}_2)\psi\right)\nonumber 
\\
\mathcal{L}_{D_3}&=
\sqrt{2(1+\Omega^2)}\left(\mu\cos\alpha(Z_3-\bar{Z}_1-Z_2+\bar{Z}_2 )
  + \psi(Z_3-\bar{Z}_3)-(Z_2-\bar{Z}_2)\psi\right)\nonumber 
\end{align}}

In this case, \eqref{S4} becomes an action for 5 complex matrices. 
For the partition function we need the independent product of 
two copies of \eqref{S4}, i.e.\ we are dealing with a complex
10-matrix model. This is already cumbersome and shows that there is
little hope to treat the original model (\ref{SpecAct}) where the 
simplifying consequences of finite matrices are not available.

The next step is to define the estimator for the average values of
interest and to specify some numerical parameters in order to analyse
the numerical results.

\section{Definition of the observables}

Following Monte Carlo methods, we will produce a sequence of
configurations $\{(\psi,Z_i)_j \}_{j = 1, 2,\cdots,T_{MC}}$ and
evaluate the average of the observables over that set of
configurations. These sequences of configurations, called Monte Carlo
chain, are representatives of the configuration space at given
parameters.  In this framework the expectation value is approximated as
\begin{equation}
\langle O\rangle \approx \frac{1}{T_{MC}}\sum_{j=1}^{T_{MC}}O_j\;,
\end{equation}
where $O_j$ is the value of the observable $O$ evaluated in the
$j$-sampled configuration, $(\psi,Z_i)_j$, $O_j= O[(\psi,Z_i)_j]$.
The internal energy is defined as
\begin{equation}
E(\Omega,\mu,\alpha)= \langle S\rangle\;,
\end{equation}
and the specific heat takes the form
\begin{equation}
C(\Omega,\mu,\alpha)= \langle S^2\rangle - \langle S\rangle^2\;.
\end{equation}
These quantities correspond to the usual definitions for energy
\begin{equation}
E(\Omega,\mu,\alpha) = 
-\frac{1}{\mathcal{Z}}\frac{\partial\mathcal{Z}}{\partial\beta}
\end{equation}
and specific heat
\begin{equation}
C(\Omega,\mu,\alpha) =\frac{\partial E}{\partial\beta}\;,
\end{equation} 
where $\mathcal{Z}$ is the partition function.  It is very useful to
compute separately the average values of the four contributions:
\begin{align}
F(\Omega,\mu,\alpha)&= \langle \operatorname{Tr} \mathcal{L}_F \rangle \;,
\\
V_0(\Omega,\mu,\alpha) &= \langle \operatorname{Tr} \mathcal{L}_{V_0}
\rangle \;,
\\
V_1(\Omega,\mu,\alpha) &= \langle \operatorname{Tr} \mathcal{L}_{V_1}
\rangle \;,
\\
D(\Omega,\mu,\alpha)&= \langle
\operatorname{Tr}\left(\mathcal{L}_{D_0}\bar{\mathcal{L}}_{D_0} 
+\dots +\mathcal{L}_{D_3}\bar{\mathcal{L}}_{D_3} 
\right) \rangle \;.
\end{align}

\subsection{Order parameters}
The previous quantities are not enough if we want to measure the
various contributions of different modes of the fields to the
configuration $(\psi,Z_i)$. Therefore, we need some control parameters
usually called order parameters. As a first idea we can think about a
quantity related to the norms of the fields, for example the sums
$\sum_{nm} |\psi_{nm}|^2$, $\sum_{nm} |Z_{inm}|^2$. These quantities are
called the full-power-of-the-field \cite{order-par,order-par1}; they
can be computed as the trace of the square:
\begin{align}
\varphi^2_a &= \operatorname{Tr}(|\psi|^2) \label{vara} \\
Z^2_{ia} &= \operatorname{Tr}(|Z_i|^2) 
\end{align}
In contrast, $\langle\varphi_a\rangle$ alone is not a good order
parameter because it does not distinguish contributions from the
different modes. But we can use it as a reference to define the
quantities
\begin{align}
\varphi^2_0 &= \sum^N_{n=0} |a_{nn}|^2 \;, \nonumber \\
Z^2_{i0} &= \sum^N_{n=0} |z_{inn}|^2\;, \label{var0}
\end{align}
where $a_{mn}$ and $z_{imn}$ are the expansion coefficients of $\psi$
and $Z_i$, respectively, in the matrix base \eqref{mb} .  Referring to
\eqref{mb} it is easy to see that these parameters \eqref{var0} are
connected with the purely spherical contribution.  These quantities
will be used to analyse the spherical contribution to the
full-power-of-the-field.  We can generalise the previous quantity and
define parameters $\varphi_l$ in such a way that they form a
decomposition of the full-power-of-the-fields:
\begin{equation}
\varphi^2_a=\varphi^2_0+\sum_{l>0} \varphi^2_l , \qquad  
Z_{ia}^2=Z_{i0}^2+\sum_{l>0} Z_{il}^2\;.
\end{equation}
Following this prescription, the other quantities for $l>0$ can be  defined as:
\begin{equation}
\varphi^2_l =\sum^l_{n,m=0} |a_{nm}(1-\delta_{nm})|^2 , \qquad  
Z_{il}^2 =\sum^l_{n,m=0} |z_{lnm}(1-\delta_{nm})|^2  \;.\label{varl}
\end{equation}
If the contribution is dominated by the spherically symmetric
parameter we expect to have $\langle\varphi^2_a\rangle \sim
\langle\varphi^2_0\rangle$, $\langle Z_{ia}^2\rangle \sim \langle
Z^2_{i0}\rangle$. 

In the next simulations we will evaluate, apart from $l =0$, the
quantity with $l = 1$ as representatives of those contributions where
the rotational symmetry is broken.  
According to  \eqref{varl} we have
\begin{equation}
\varphi^2_1 =|a_{10}|^2+ |a_{01}|^2, \  Z_{i1}^2 =|z_{i10}|^2+ |z_{i01}|^2\;.
\end{equation}
Using higher $l$ in \eqref{varl}
we could analyse the contributions of the remaining modes, but it turns out
that the measurements of the first two modes are enough to
characterise the behaviour of the system.

\section{Numerical results }

Now we discuss the results of the Monte Carlo simulation of the
approximated spectral model. As a first approach we use some
restrictions on the parameters.  Starting point is the approximation
\eqref{S4} of the spectral action. Since \eqref{S4} is symmetric under
the transformation $\mu\mapsto-\mu$ we can assume $\mu\geq0 $ and
$\mu^2\geq 0$. In this first treatment we explore the range
$\mu\in[0,3.1]$, which is enough to show a particular behaviour of the
system for fixed $\Omega$. The parameter $\Omega$ appears only with
its square and is defined as a real parameter, therefore also for
$\Omega$ we require $\Omega\geq0$.  For the scalar model
\cite{Grosse:2004yu} it was possible to restrict to $\Omega \in
[0,1]$, because Langmann-Szabo duality maps $\Omega$ to
$\frac{1}{\Omega}$.  In the gauge model under consideration,
Langmann-Szabo duality is not realised.  Due to the prefactor in front
of the integral \eqref{S4}, the action vanishes for $\Omega \to
\infty$.  Studying the plots for the energy and the specific heat, we
have chosen the range $\Omega \in [0,2\pi]$ in which the action is
significantly different from zero.  The last parameter to consider is
$\alpha$, which is connected to the choice of the vacuum state, with range
 $\alpha \in [0,2\pi]$. The study of the system varying $\alpha$
is quite important from a theoretical point of view because
it is related to the vacuum invariance.  In the action there appear
some contributions proportional to $(\sin\alpha)/\Omega$ which seem to
diverge for $\Omega=0$. Numerically we have verified that this is an
eliminable divergence and the curves of the observables can be
extended to $\Omega=0$ by continuity.  Studying the dependence on $\alpha$
we can conclude that in the limit $N\to\infty$ the observables are
independent from $\alpha$, therefore for our purposes $\alpha$ will be
fixed equal to zero avoiding the annoying terms. In general, for each
observable we compute the plots for matrix size
approximations $N =$5, 10, 15, 20.

\subsection{Varying $\alpha$ }

We start looking at the variation of the energy density and of the 
full-power-of-the-fields density for fixed $\mu$ and $\Omega$, varying $\alpha
\in [0,2\pi]$. As representatives we present the plots for
$\mu=1$, $\Omega\in\{1,\,0.5\}$, but we obtain the same behaviour for any
other choice of the parameters allowed in the considered range.
\begin{figure}[htb]
\begin{center}
\includegraphics[scale=0.55]{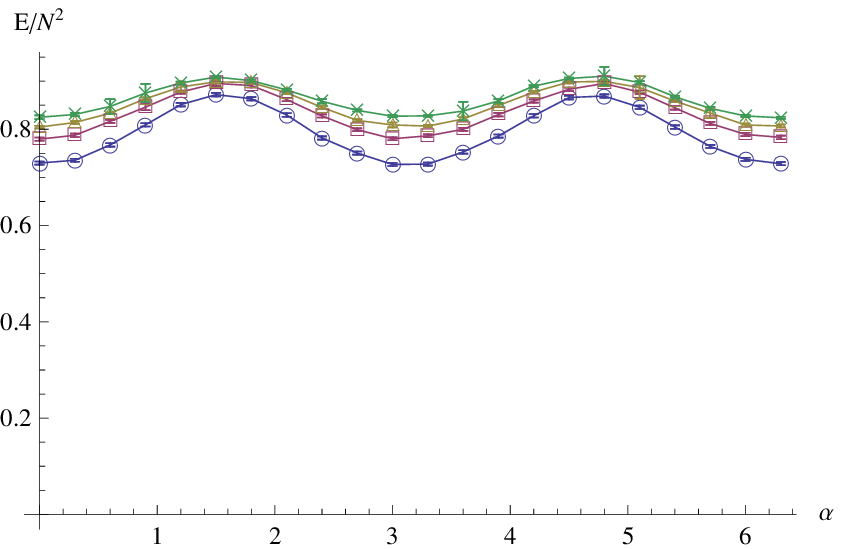}
\includegraphics[scale=0.55]{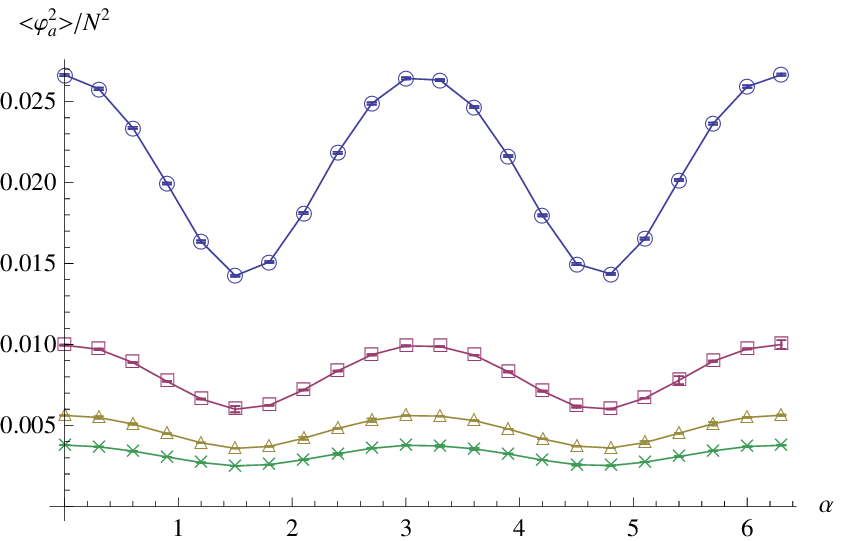}
\includegraphics[scale=0.55]{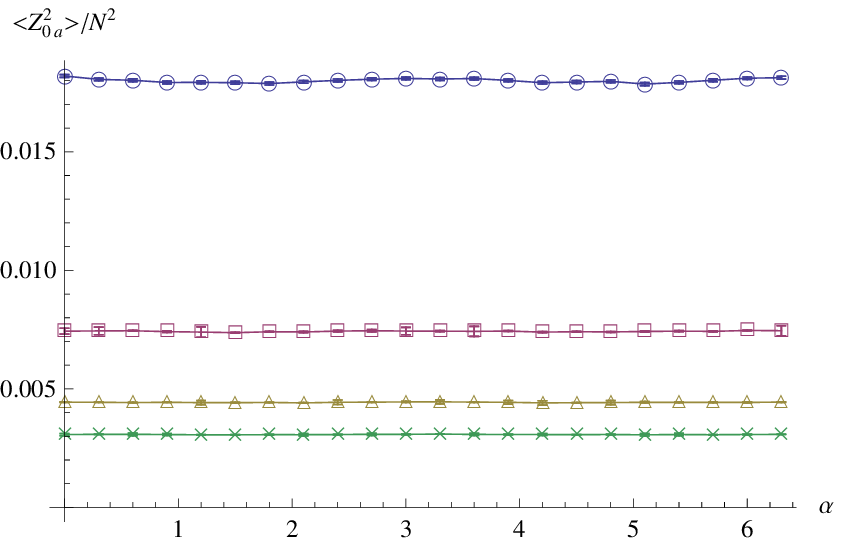}
\end{center}
\caption{\footnotesize Total energy density and
  full-power-of-the-fields density for $\langle\varphi^2_a\rangle$,
  $\langle Z_{0a}^2 \rangle$ (from the left to the right) fixing
  $\mu=1$, $\Omega=1$, varying $\alpha$ and $N$.  $N=5$ (circle),
  $N=10$ (square), $N=15$ (triangle), $N=20$ (cross).
  \normalsize}\label{Figure 1}\end{figure}
All three plots show an oscillating behaviour of the values, and this
oscillation is present in all other quantities measured. The amplitude
of this oscillation becomes smaller and smaller increasing the size of
the matrix and this is true for all the quantities. The same trend is
described in fig.\ref{Figure 2} which shows different positions of 
the maxima, but again smaller amplitudes for increasing $N$.
\begin{figure}[htb]
\begin{center}
\includegraphics[scale=0.55]{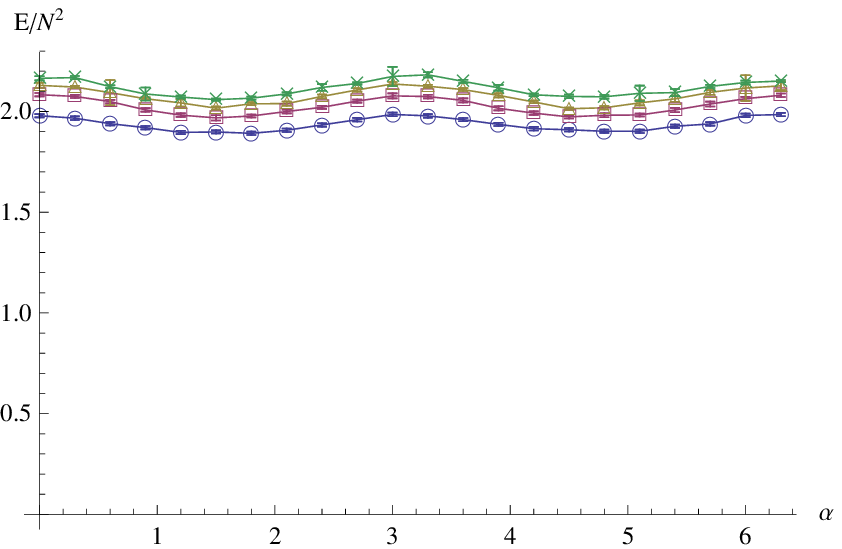}
\includegraphics[scale=0.55]{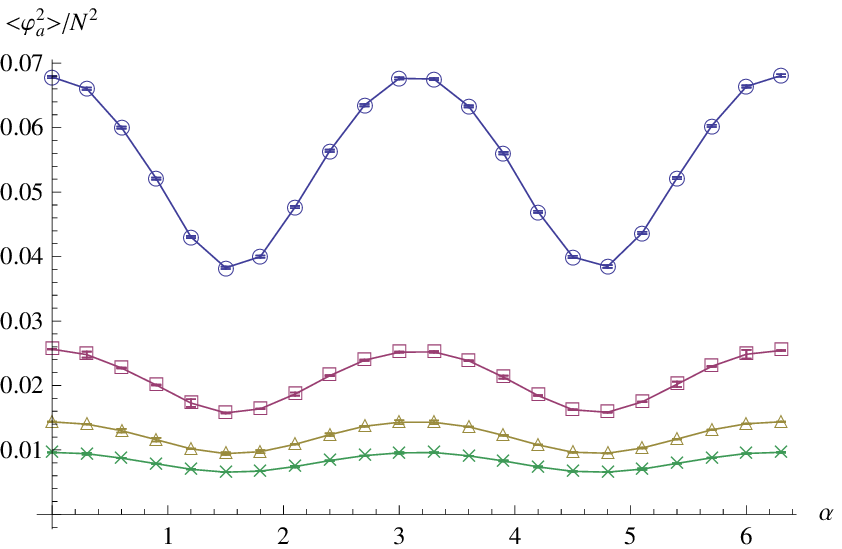}
\includegraphics[scale=0.55]{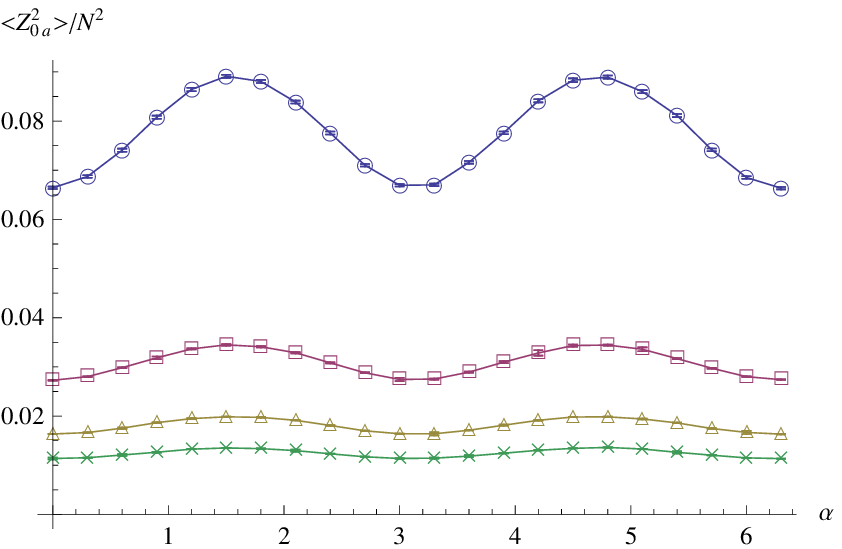}
\end{center}
\caption{\footnotesize Total energy density and
  full-power-of-the-fields density for $\langle\varphi^2_a\rangle$,
  $\langle Z_{0a}^2 \rangle$ (from the left to the right) fixing
  $\mu=1$, $\Omega=0.5$, varying $\alpha$ and $N$.
  \normalsize}\label{Figure 2}\end{figure}
These results allow us to consider $\alpha=0$ for all next plots,
since we are interested in the behaviour of the system for $N \to
\infty $. This occurrence simplify all the next simulations thanks to
the vanishing of terms $\sim(\sin\alpha)/\Omega $ appearing in the
discretised action. 

\subsection{Varying $\Omega$ }

\begin{figure}[htb]
\begin{center}
\includegraphics[scale=0.55]{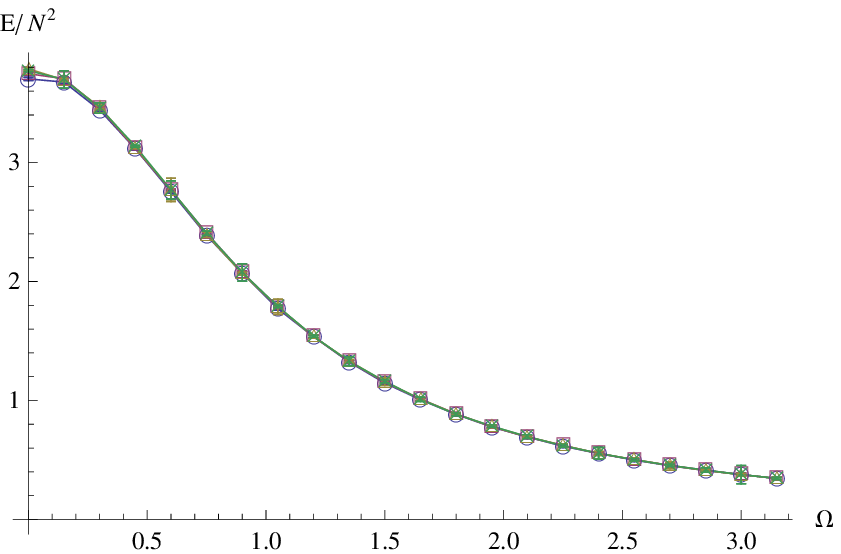}
\includegraphics[scale=0.55]{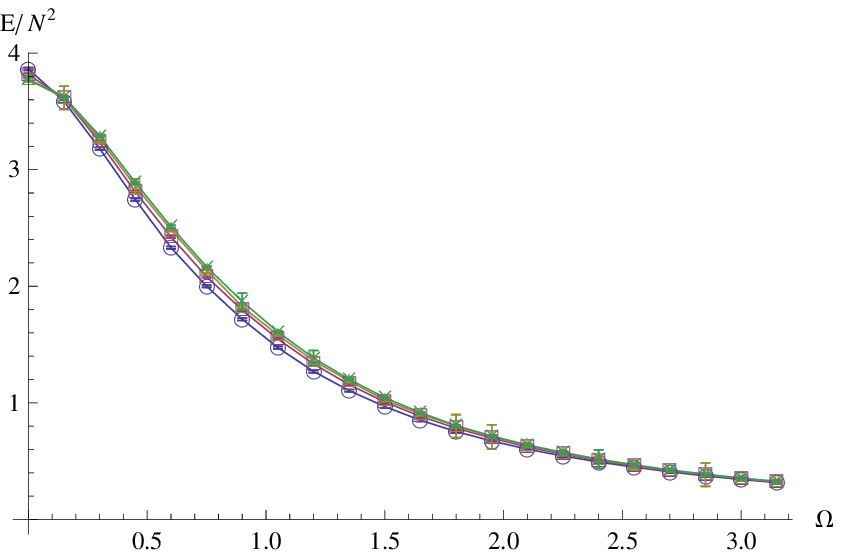}
\end{center}
\caption{\footnotesize Total energy density and the various
  contributions for $\mu=0$ (left), $\mu=0$ (right), $\alpha=0$ varying
  $\Omega$ and $N$. With $N=5$ (circle), $N=10$ (square), $N=15$
  (triangle), $N=20$ (cross).\normalsize}\label{Figure 2-bis}
\end{figure}

As already mentioned we chose $[0,3]$ as range for $\Omega$. In fact, if
we look at the plots in fig.\ref{Figure 2-bis} of the total energy
density $\langle S \rangle / N^2$ for $\mu\in \{0,\,1\}$, we notice that the
action tends to zero for $\Omega$ outside
the selected interval. This behaviour of the action is the same for
all possible choices of parameters and for the specific heat, too.

\begin{figure}[htb]
\begin{center}
\includegraphics[scale=0.55]{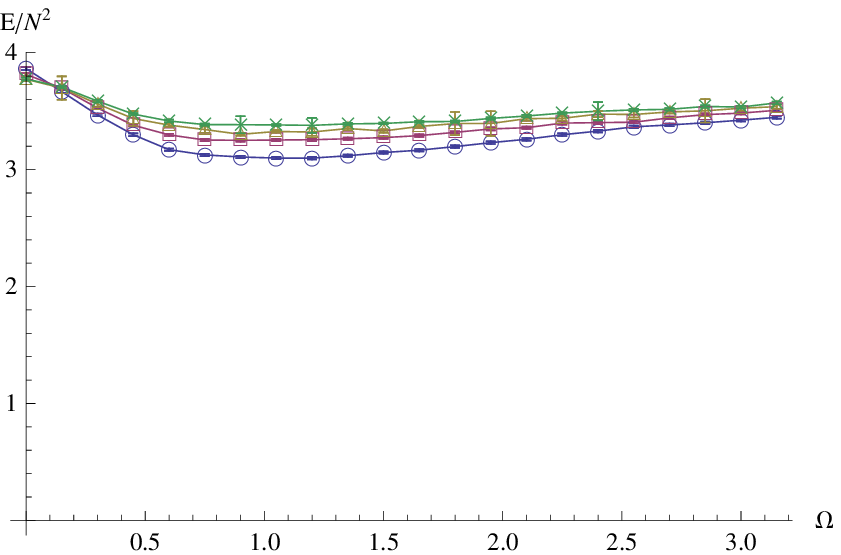}
\includegraphics[scale=0.55]{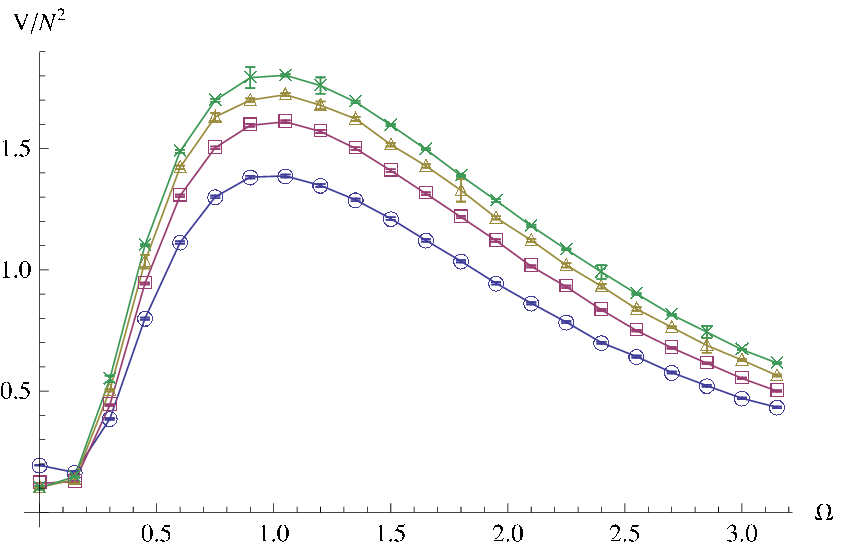}
\includegraphics[scale=0.55]{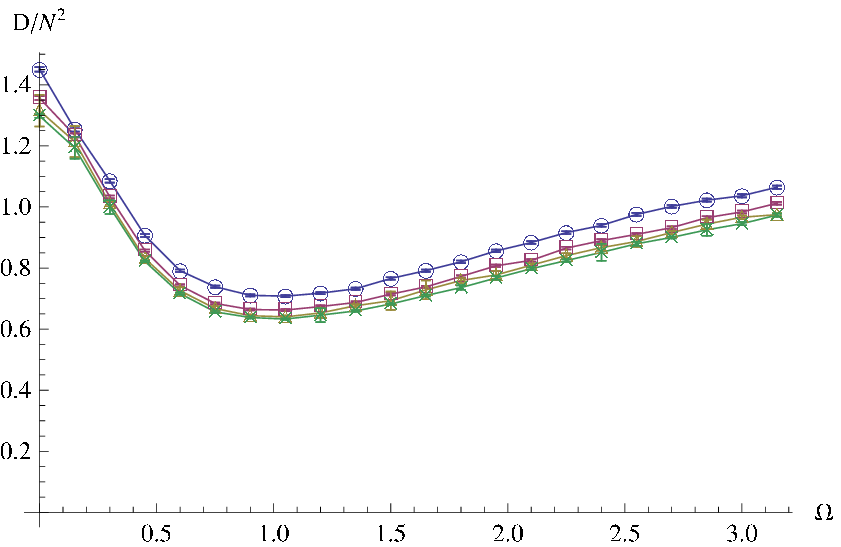}
\includegraphics[scale=0.55]{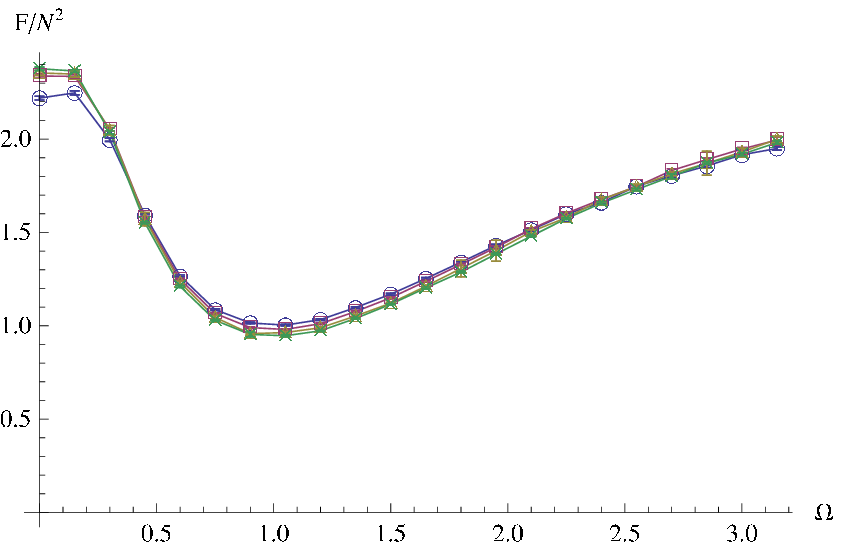}
\end{center}
\caption{\footnotesize Total energy density and the various
  contributions for $\mu=1$, $\alpha=0$ varying $\Omega$ and $N$. From
  the left to the right $E$, $V$, $D$, $F$ with $N=5$ (circle), $N=10$
  (square), $N=15$ (triangle), $N=20$
  (cross).\normalsize}\label{Figure 4}\end{figure}

In the rest of this section we ignore for the computations of $\langle
E\rangle,\langle D\rangle,\langle V\rangle,\langle F\rangle$ the
global prefactor $(1 +\Omega^2)^{-1}$.  In this way we focus our
attention to the integral as the source of possible phase transitions.
Now we will analyse three cases in which $\mu$ is fixed to $0,1,3$. In
all cases $\alpha$ is zero and we vary $\Omega\in [0,3]$.  The plots
in fig.\ref{Figure 4} show the total energy density and the various
contributions: the potential $V / N^2$, the Yang-Mills part $F / N^2$
and the covariant derivative part $D / N^2$, for $\mu=1$. There is no
evident discontinuity or peak, and increasing the size of the matrices
the curves remain smooth.

\begin{figure}[htb]
\begin{center}
\includegraphics[scale=0.7]{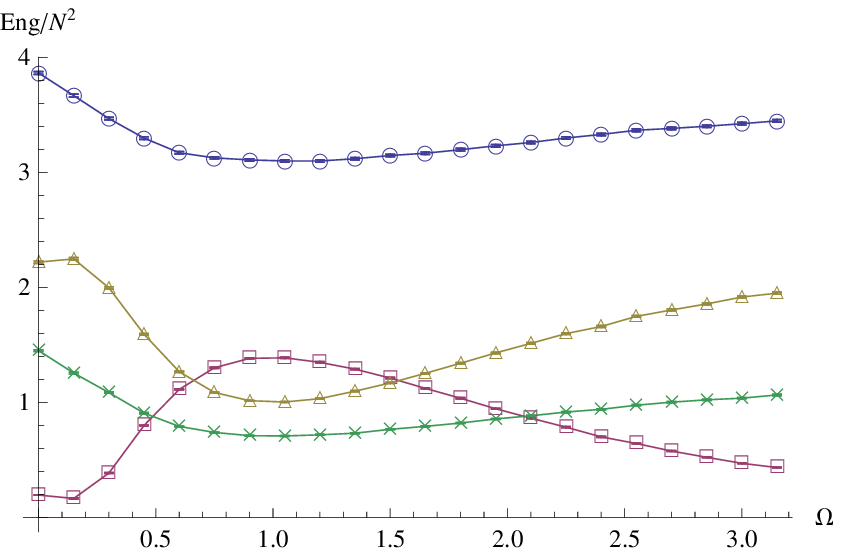}
\includegraphics[scale=0.7]{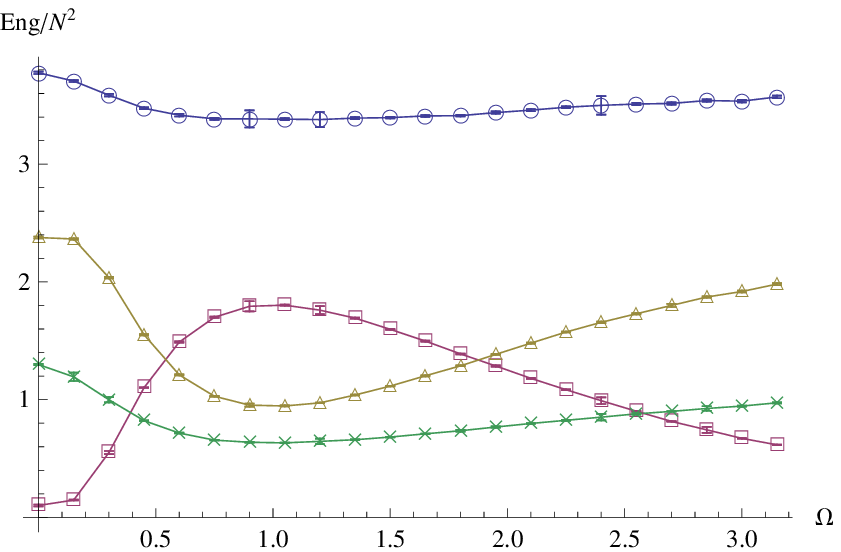}
\end{center}
\caption{\footnotesize Comparison of the total energy density and the
  various contributions for $\mu=1$, $\alpha=0$. $E$ (circle), $F$
  (triangle), $D$ (cross), $V$ (square). With $N=5$ (left) and $N=20$
  (right).  \normalsize}\label{Figure 5}
\end{figure}

Comparing the energy density and the various contributions
in fig.\ref{Figure 5} we notice that the contributions between $F$ and
$V$ balance each other and the total energy follows the slope of $D$,
and this  behaviour continues increasing the size of the matrices.

\begin{figure}[htb]
\begin{center}
\includegraphics[scale=0.7]{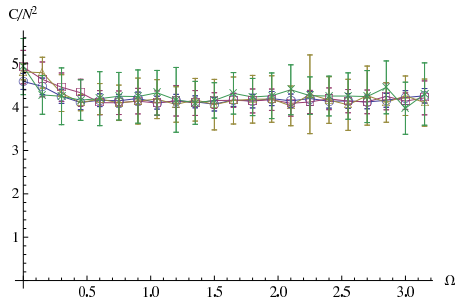}
\end{center}
\caption{\footnotesize Specific heat for $\mu=1$.\normalsize}\label{Figure 6}
\end{figure}

The specific heat density in fig.\ref{Figure 6} shows a small peak in
$\Omega=0$. This peak does not increases as $N$ increases, therefore is
not related to a phase transition.

\begin{figure}[htb]
\begin{center}
\includegraphics[scale=0.75]{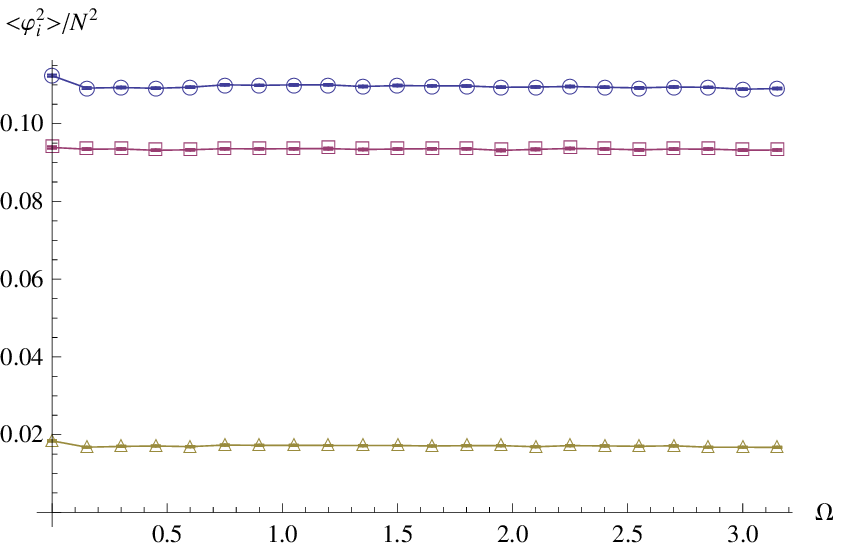}
\includegraphics[scale=0.55]{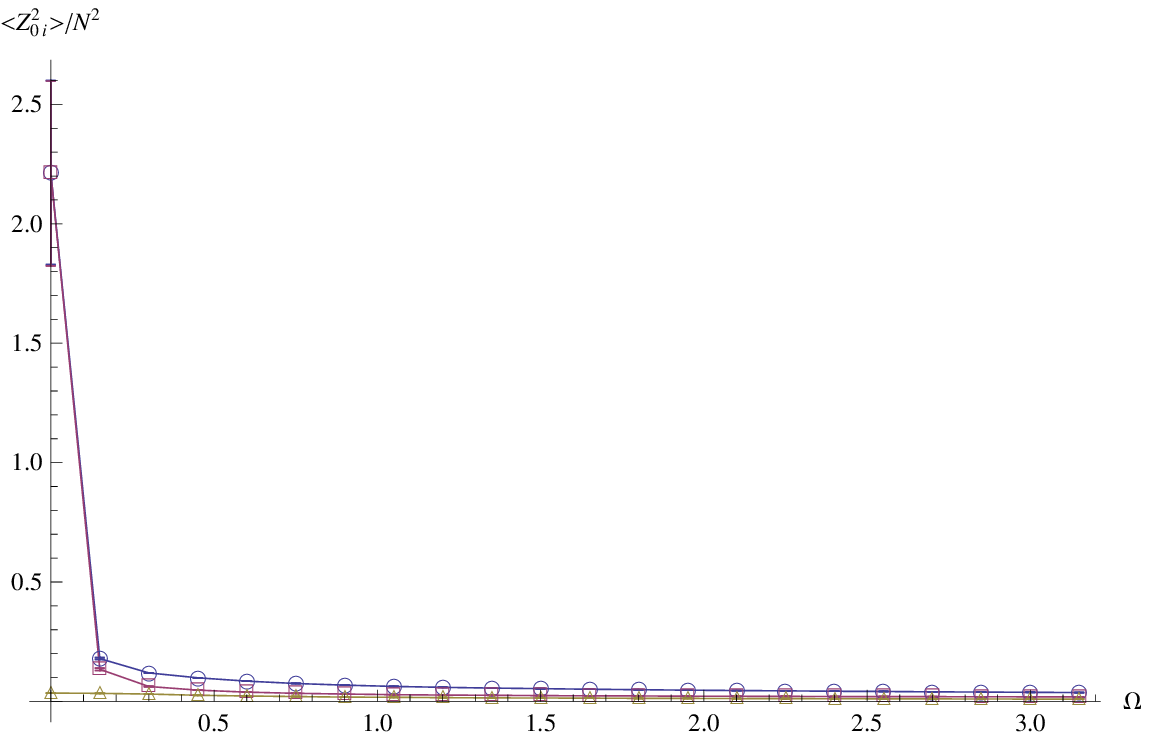}
\end{center}
\caption{\footnotesize On the left comparison of
  $\langle\varphi_a^2\rangle $ (circle), $\langle\varphi^2_0 \rangle$
  (square) and $\langle\varphi^2_1\rangle $ (triangle) density. On the
  right comparison of $\langle Z_{0a}^2 \rangle$ (circle), $\langle
  Z_{00}^2 \rangle$ (square) and $\langle Z_{01}^2\rangle$ (triangle)
  density.\normalsize}\label{Figure 8}
\end{figure}

\begin{figure}[htb]
\begin{center}
\includegraphics[scale=0.55]{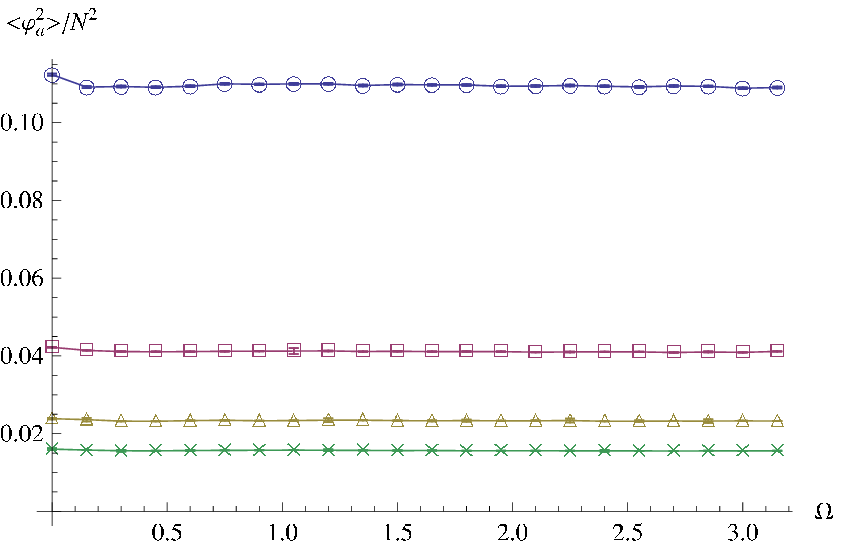}
\includegraphics[scale=0.55]{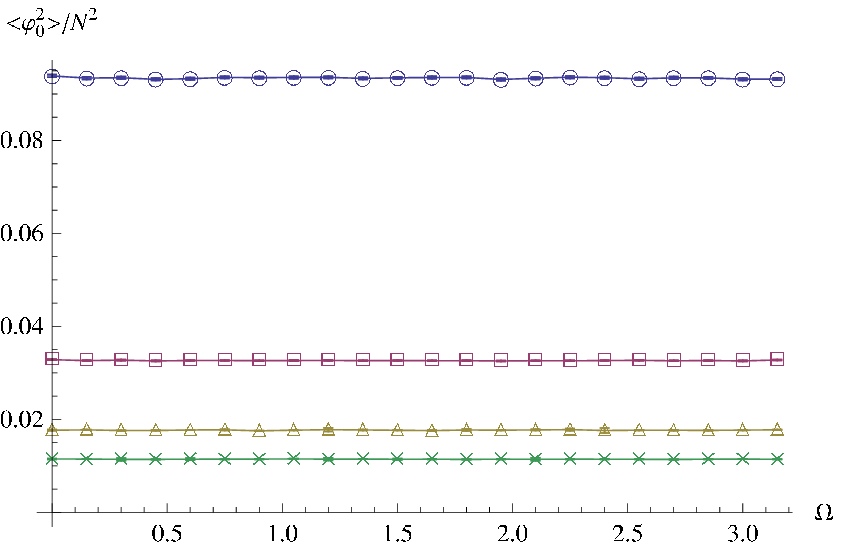}
\includegraphics[scale=0.55]{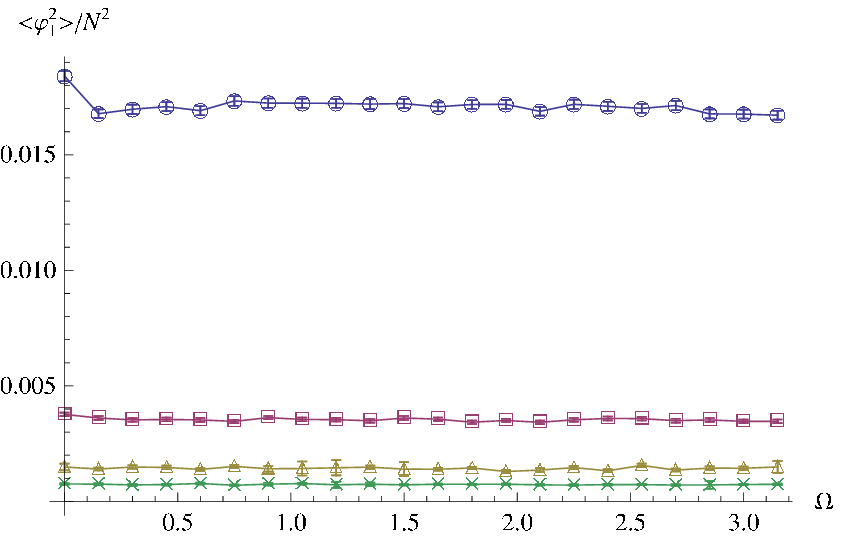}
\includegraphics[scale=0.55]{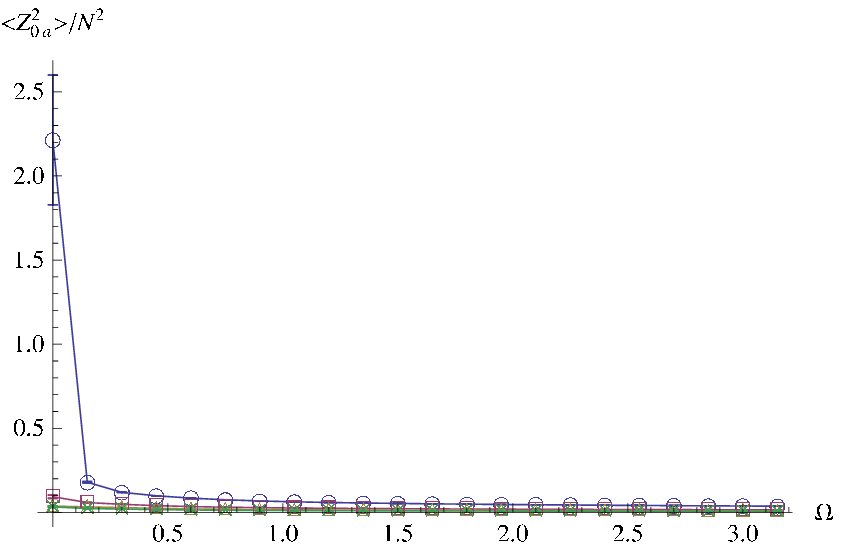}
\includegraphics[scale=0.55]{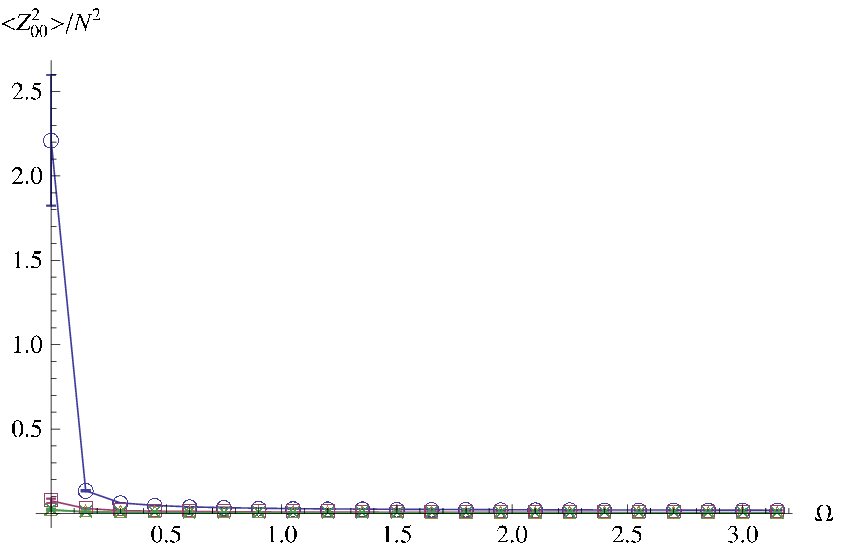}
\includegraphics[scale=0.55]{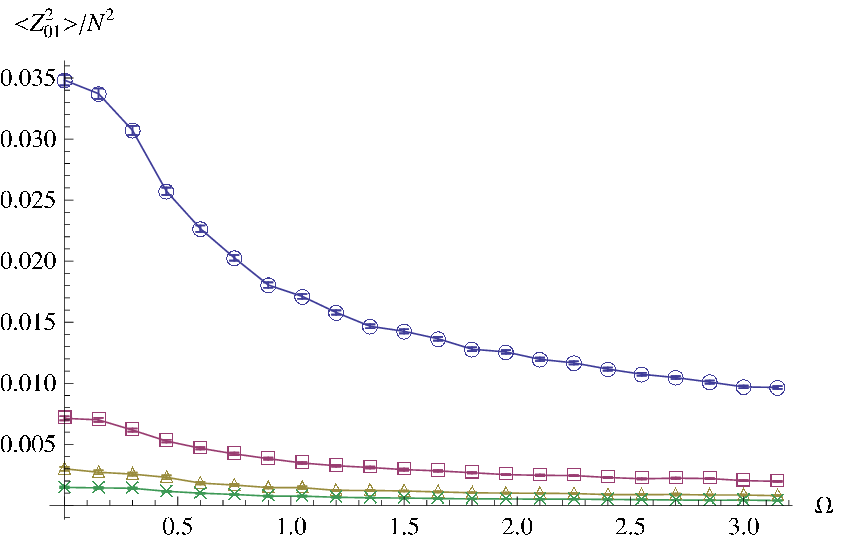}
\end{center}
\caption{\footnotesize Starting from the up left corner and from the
  left to the right the densities for $\langle\varphi_a^2\rangle $, $\langle\varphi^2_0
  \rangle$, $\langle\varphi^2_1\rangle $, $Z_{0a}^2$, $\langle Z_{00}^2 \rangle $ and $\langle Z_{01}^2\rangle$
  for $\mu=1$ varying $\Omega$ and $N$. \normalsize}\label{Figure 9}
\end{figure}

In order to gain some information on the composition of the fields we
look at the order parameters defined in the previous section. Starting
from the scalar field $\psi$, fig.\ref{Figure 8} shows the plots for
$\langle\varphi_a^2\rangle $, $\langle\varphi^2_0 \rangle$ and
$\langle\varphi^2_1\rangle $ for $N=5$.  The three values
$\langle\varphi_a^2\rangle $, $\langle\varphi^2_0 \rangle$ and
$\langle\varphi^2_1\rangle$ seem essentially constant, where the
spherical contribution $\langle\varphi^2_0 \rangle$ to the
full-power-of-the-field is dominant.
The behaviour of the $Z_0$ fields (which describe the covariuant
coordinates) is different. Here the spherical contribution becomes
dominant only for $\Omega$ approaching $0$, starting from a zone in
which the contribution of $\langle Z_{00}^2 \rangle$ and $\langle
Z_{01}^2\rangle $ are comparable.  For brevity we only show the plots
for $\langle Z_{0a}^2 \rangle $, $\langle Z_{00}^2 \rangle$ and
$\langle Z_{01}^2 \rangle$, but taking into account the statistical
errors, the other $Z_i$-related plots are compatible to the
$Z_0$-case. The dependence of the previous quantities on $N$ is shown
in the following plots fig.\ref{Figure 9}.
All previous parameters decrease with $N$, but the dominance of
$\varphi_0$ on the total-power-of-the-field is independent by $N$. The
peak related to $Z_0$ decreases with $N$, but if we look at the single 
plot for the spherical contribution at $N=20$, 
the peak persists as $\Omega$ approaches $\Omega=0$.

\begin{figure}[htb]
\begin{center}
\includegraphics[scale=0.55]{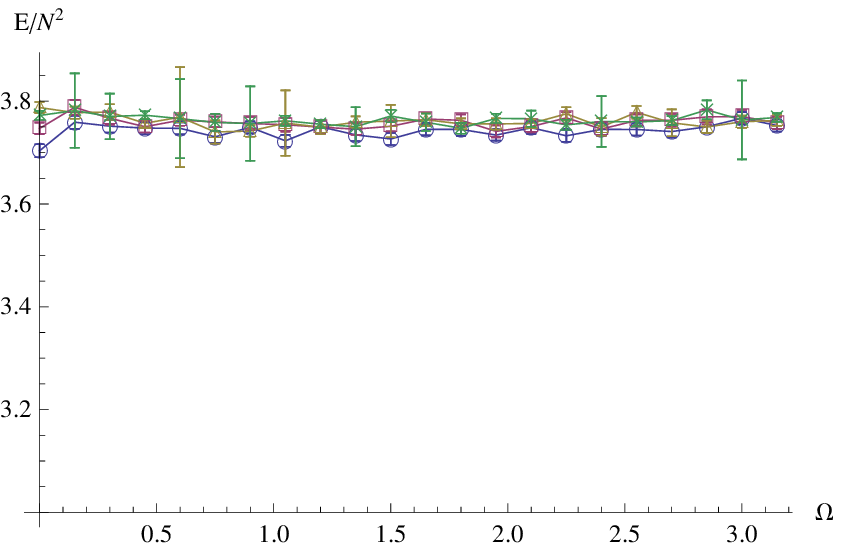}
\includegraphics[scale=0.55]{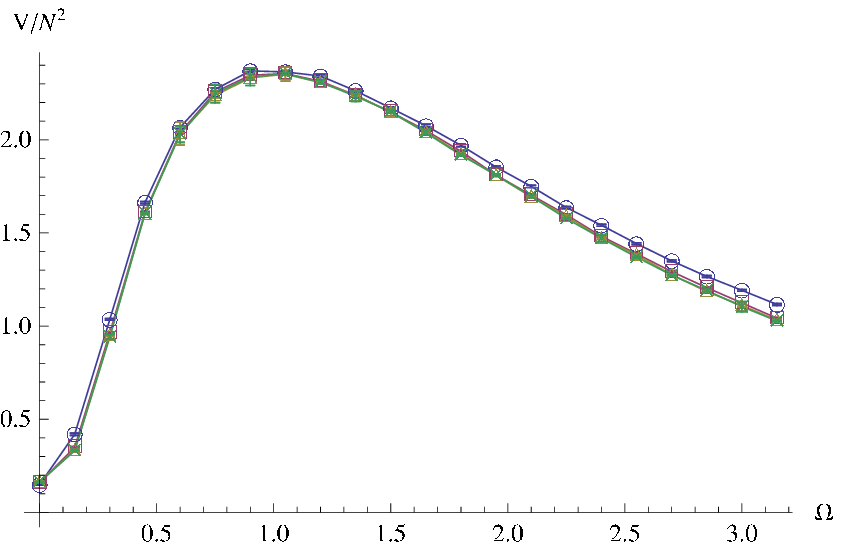}
\includegraphics[scale=0.55]{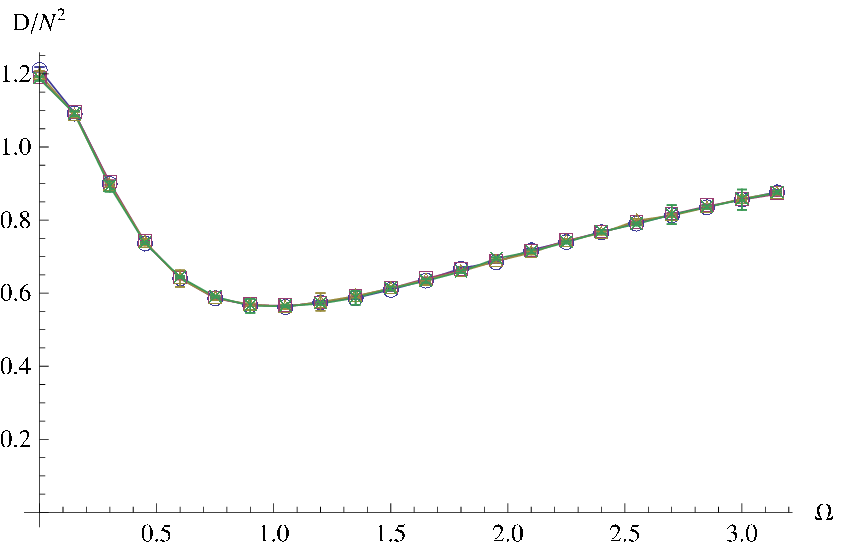}
\includegraphics[scale=0.55]{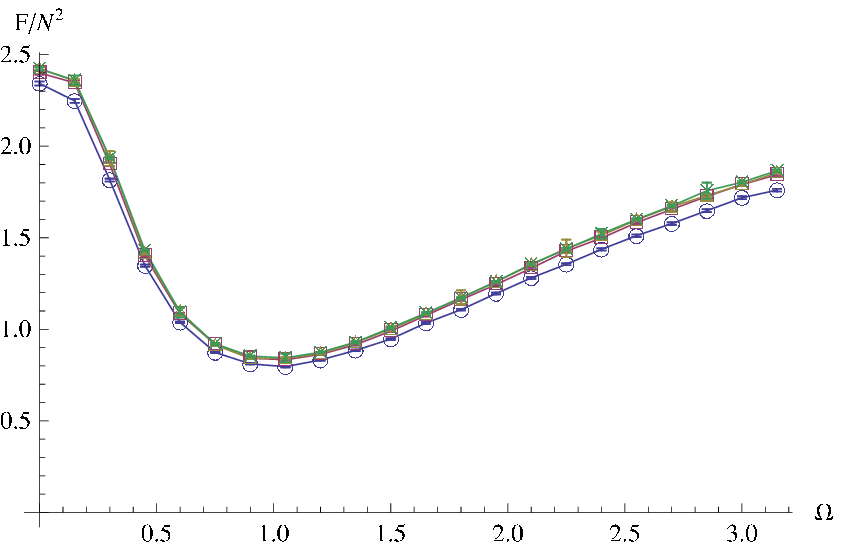}
\includegraphics[scale=0.55]{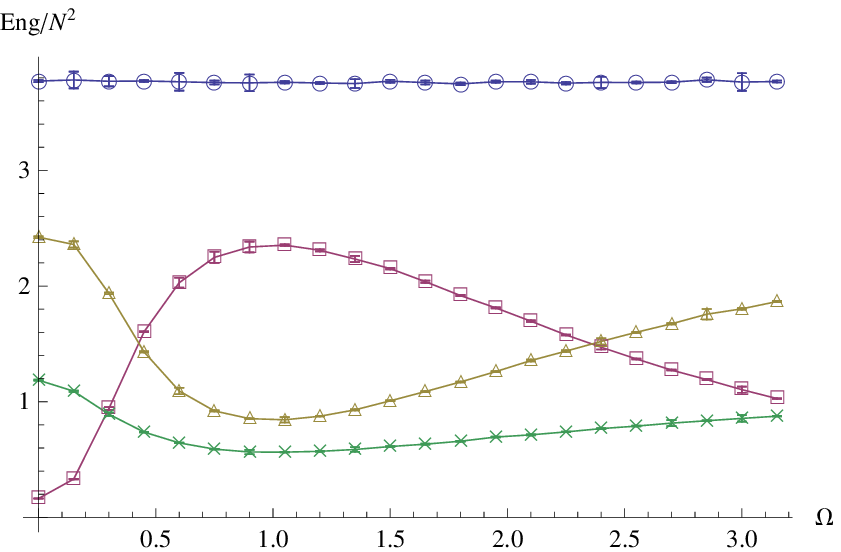}
\end{center}
\caption{\footnotesize Total energy density, various contributions and
  the comparison among them for $\mu=0$ varying $\Omega$ and $N$. From
  the left to the right $E$, $V$, $D$, $F$ and
  comparison.\normalsize}\label{Figure 10}\end{figure}

Now we will analyse the model for $\mu=0$. Fig.\ref{Figure 10} shows
the plots for total energy density and the contributions $V$, $D$,
$F$.  The slope of the total energy density seems to be constant. The
$D$-contribution and the $F$-contribution do not balance each other
like in the previous case, but all three contributions balance
themselves to produce a constant sum.

\begin{figure}[htb]
\begin{center}
\includegraphics[scale=0.7]{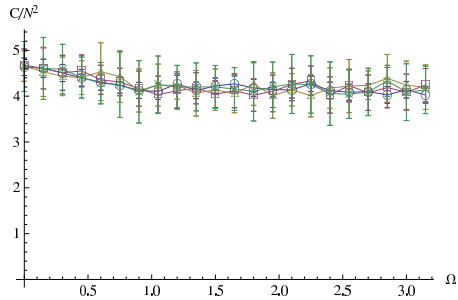}
\end{center}
\caption{\footnotesize Specific heat density for $\mu=0$ varying
  $\Omega$ and $N$.\normalsize}\label{Figure 11}
\end{figure}

The specific heat density fig.\ref{Figure 11} shows again the small peak in
$\Omega=0$ without $N$-dependence.%

\begin{figure}[htb]
\begin{center}
\includegraphics[scale=0.55]{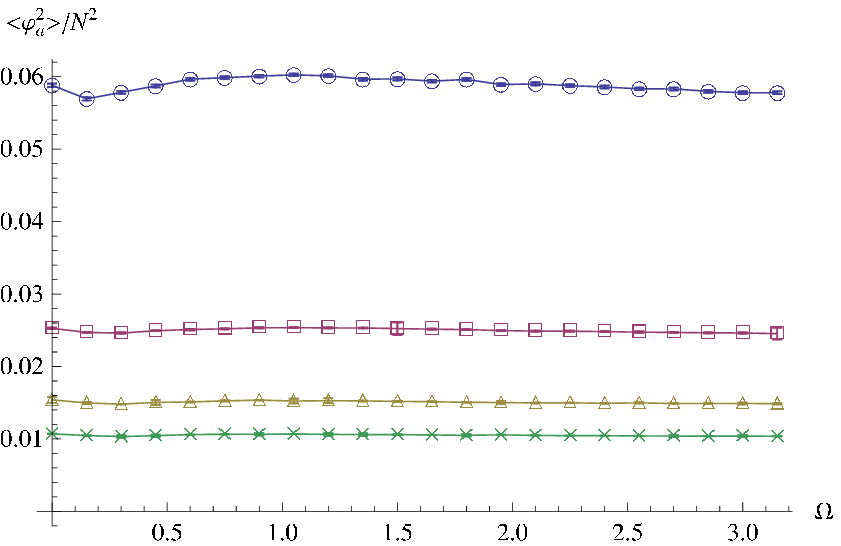}
\includegraphics[scale=0.55]{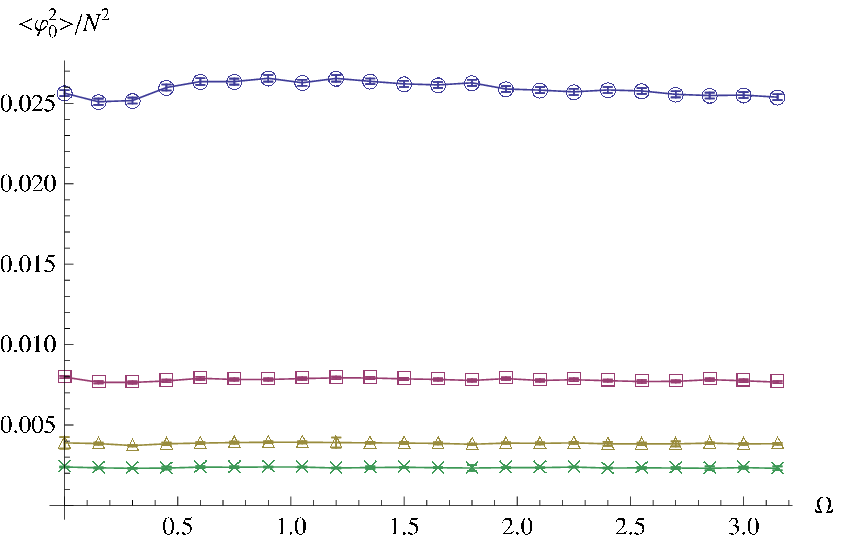}
\includegraphics[scale=0.55]{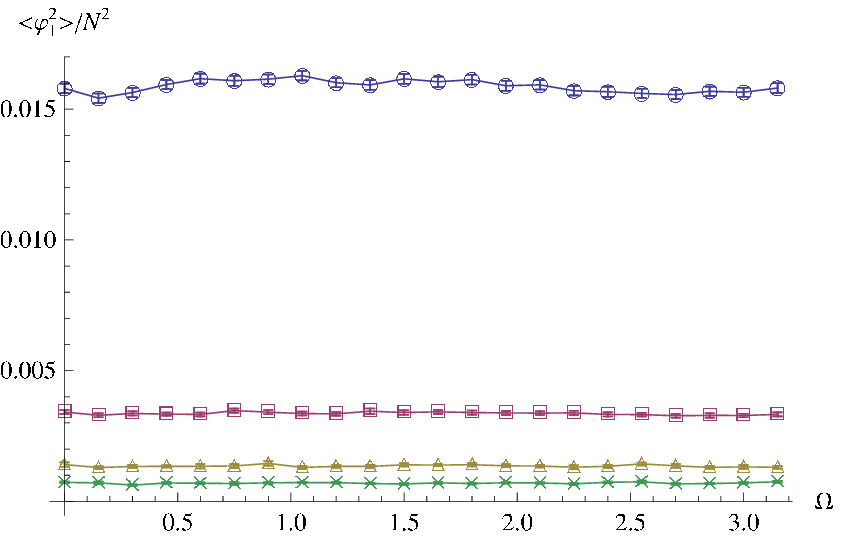}
\includegraphics[scale=0.55]{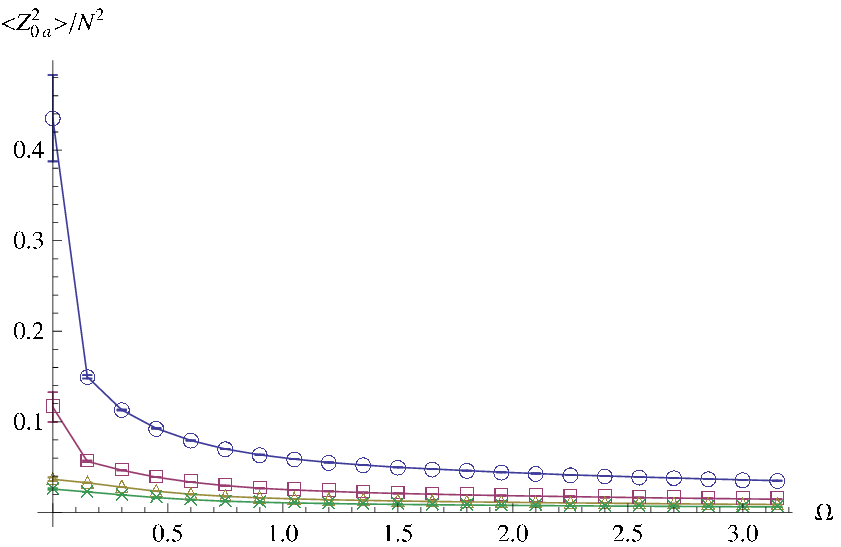}
\includegraphics[scale=0.55]{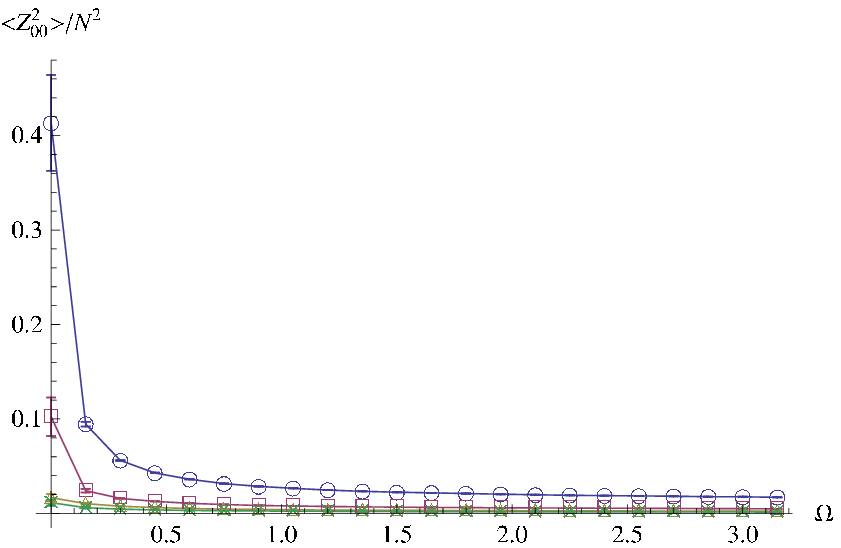}
\includegraphics[scale=0.55]{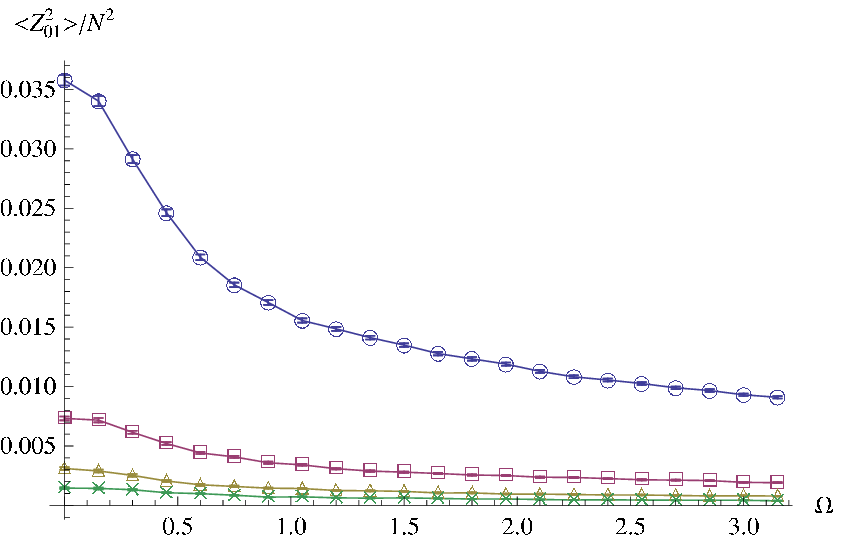}
\end{center}
\caption{\footnotesize Starting from the up left corner and from the
  left to the right the densities for $\langle\varphi_a^2\rangle $,
  $\langle\varphi^2_0 \rangle$, $\langle\varphi^2_1\rangle $, $\langle
  Z_{0a}^2 \rangle$, $\langle Z_{00}^2 \rangle$ and $\langle
  Z_{01}^2\rangle$ for $\mu=0$ varying $\Omega$ and $N$.
}\label{Figure 12}
\end{figure}

For the other quantities
$\langle\varphi_a^2\rangle $, $\langle\varphi^2_0 \rangle$, $\langle\varphi^2_1\rangle $ and $\langle Z_{0a}^2 \rangle
$, $\langle Z_{00}^2 \rangle$, $\langle Z_{01}^2\rangle$ we have according to 
fig.\ref{Figure 12} the same behaviour as in the case $\mu=1$.

\begin{figure}[htp]
\begin{center}
\includegraphics[scale=0.55]{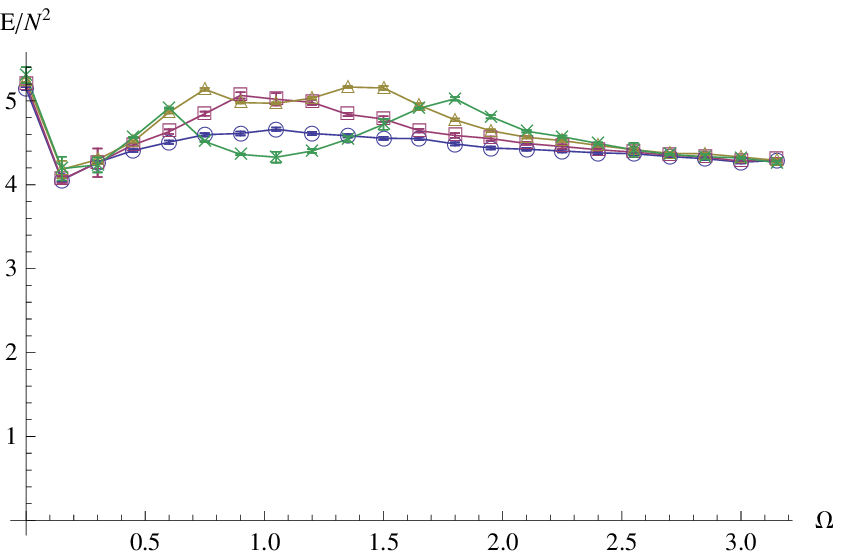}
\includegraphics[scale=0.55]{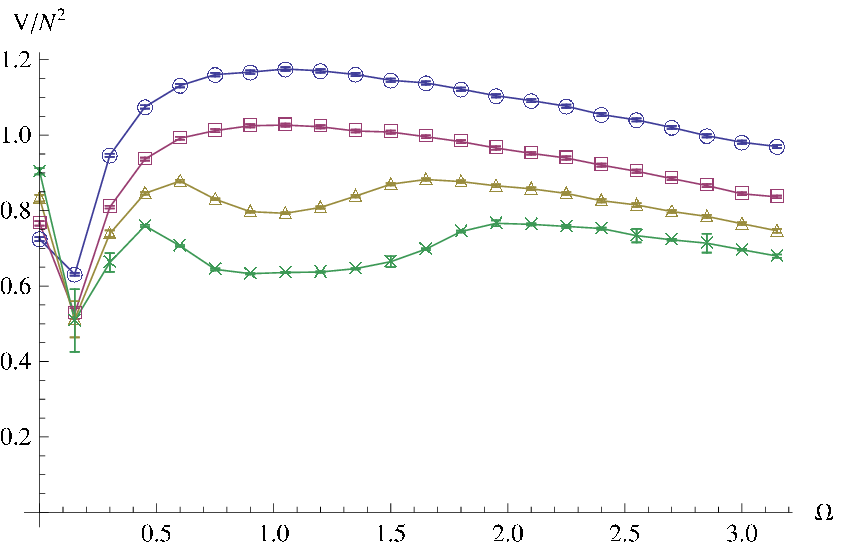}
\includegraphics[scale=0.55]{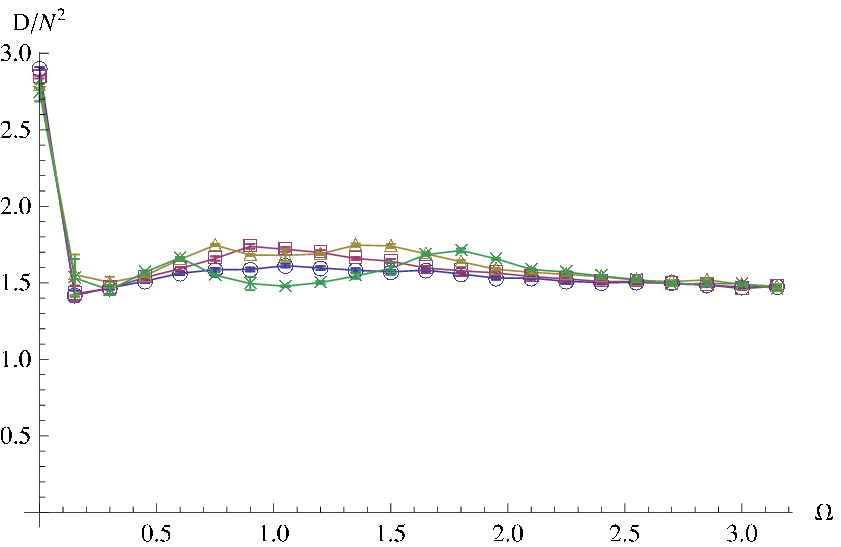}
\includegraphics[scale=0.55]{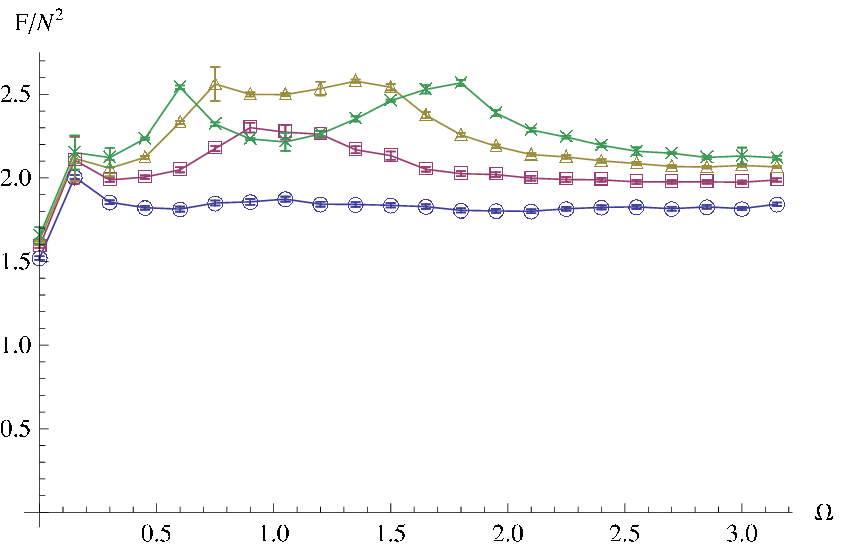}
\includegraphics[scale=0.55]{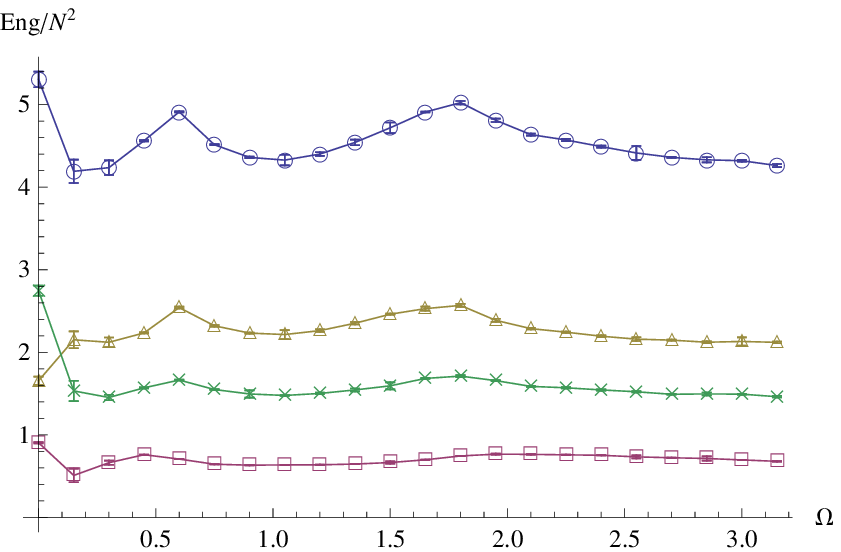}
\end{center}
\caption{\footnotesize Total energy density, various contributions and
  the comparison among them for $\mu=3$ varying $\Omega$ and $N$. From
  the left to the right  $E$, $V$, $D$, $F$ and
  comparison. \normalsize}\label{Figure 13}
\end{figure}

A completely different response of the system is obtained in the plots
for $\mu=3$, as we can see from fig.\ref{Figure 13}. 
The slope of total energy density is very similar to the $F$-component
instead of $D$. However, there appears a sharp minimum around
$\Omega=0.1$ and two maxima at $\Omega\approx 0.6$ and $\Omega \approx
1.8$ for large $N$.  This dramatic change in the plots might be
interpreted as consequence of a phase transition in the
parameter $\mu$. Actually, in the next section we will find a peak in
the specific heat density for some fixed $\Omega$ and varying
$\mu\in[0,3]$.

\begin{figure}[htp]
\begin{center}
\includegraphics[scale=.7]{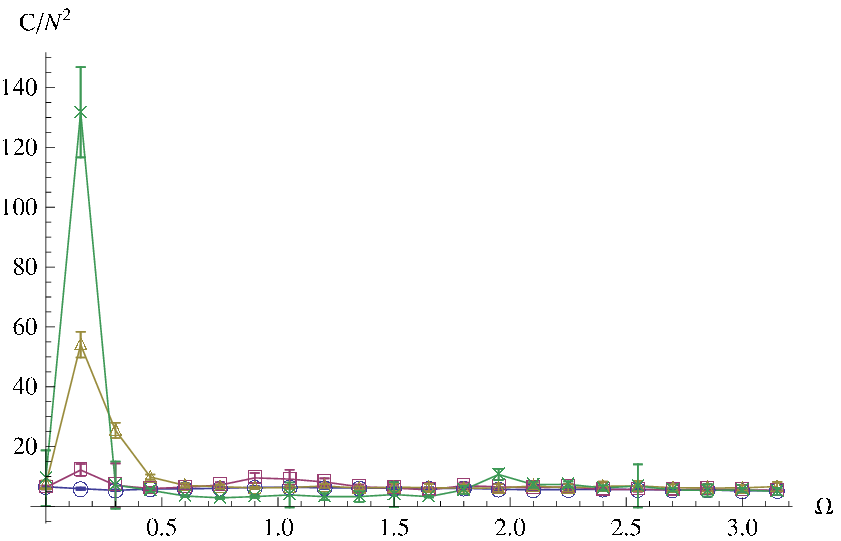}
\end{center}
\caption{\footnotesize Specific heat density for $\mu=3$ varying
  $\Omega$ and $N$.\normalsize}\label{Figure 14}
\end{figure}

The specific heat density fig.\ref{Figure 14} displays a strong
change, too. 
In fact, instead of the peak at $\Omega=0$, the peak
appears close to the origin around $\Omega=0.15$. This peak,
in contrast to the previous ones, grows as $N$ increases and therefore
could indicate a phase transition.

\begin{figure}[hbp]
\begin{center}
\includegraphics[scale=0.55]{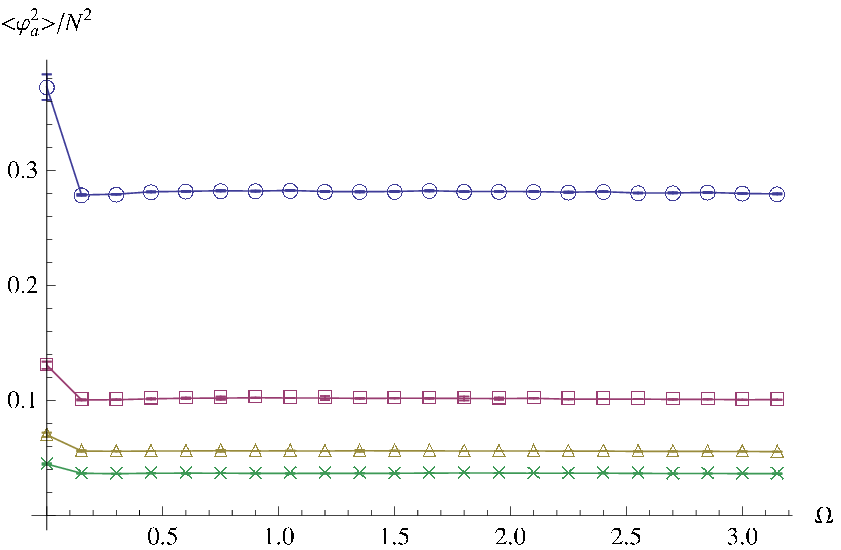}
\includegraphics[scale=0.55]{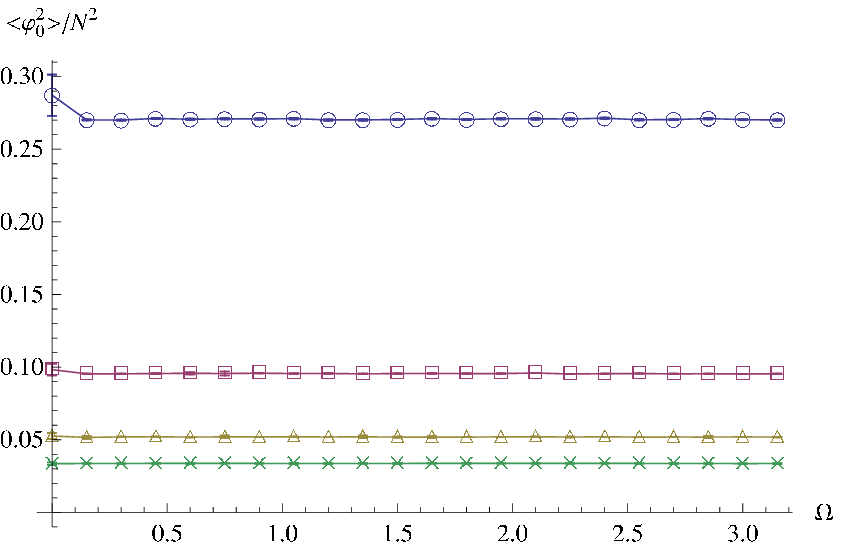}
\includegraphics[scale=0.55]{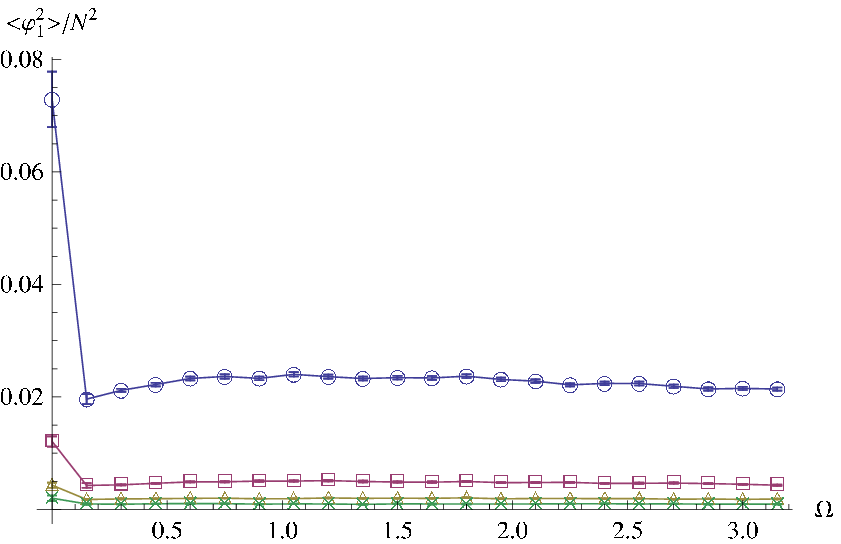}
\includegraphics[scale=0.55]{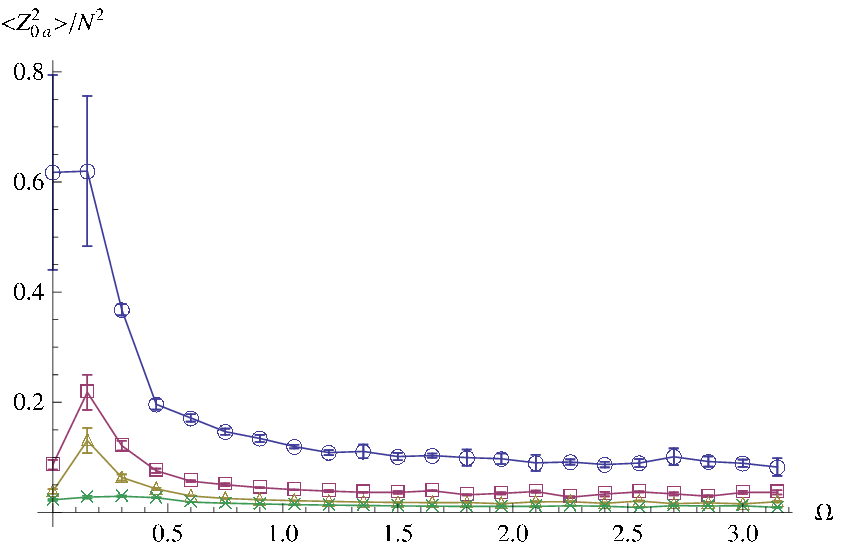}
\includegraphics[scale=0.55]{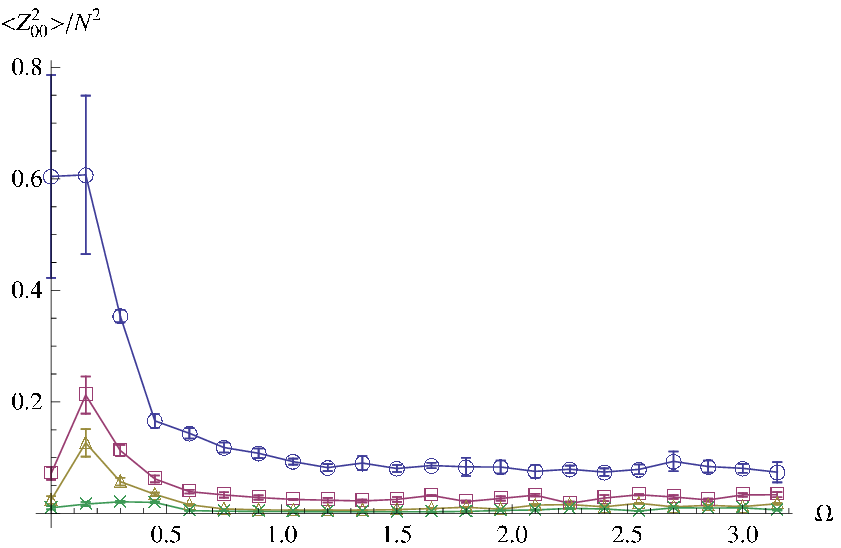}
\includegraphics[scale=0.55]{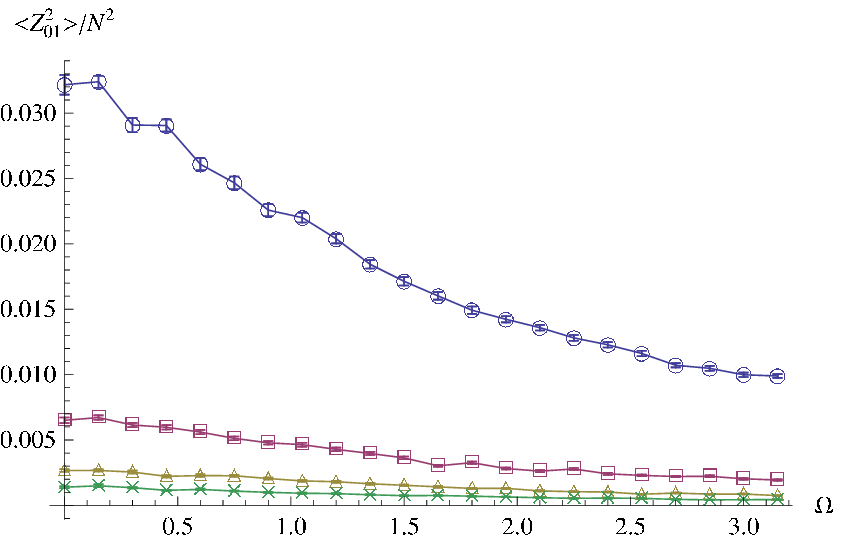}
\end{center}
\caption{\footnotesize Starting from the up left corner and from the
  left to the right the densities for $\langle\varphi_a^2\rangle $,
  $\langle\varphi^2_0 \rangle$, $\langle\varphi^2_1\rangle $, $\langle
  Z_{0a}^2 \rangle$, $\langle Z_{00}^2 \rangle $ and $\langle
  Z_{01}^2\rangle$ for $\mu=3$ varying $\Omega$ and
  $N$.\normalsize}\label{Figure 15}
\end{figure}
The fig.\ref{Figure 15} describes the behaviour of the order
parameters densities $\langle\varphi_a^2\rangle $, $\langle\varphi^2_0
\rangle$, $\langle\varphi^2_1\rangle $ and $\langle Z_{0a}^2 \rangle$,
$\langle Z_{00}^2 \rangle $, $\langle Z_{01}^2\rangle$. They show a
similar behaviour as the corresponding plots for $\mu=1$ and $\mu=0$.
For the $\psi$ field the spherical contribution remains dominant.
However, in the $\langle\varphi^2_1\rangle $ plot there appears a
deviation from the constant slope.  This deviation is evident for
$N=5$ but still present for higher $N$.  The order parameters for
$Z_0$ display a peak close to the origin without oscillations even for
$N=5$. This maximum for higher $N$ does not move closer to the origin,
in other words, this shift is not caused by finite volume effects.
Even for $Z^2_{01} $ there appears a peak at $\Omega=0$ which becomes
shifted and smoother for higher $N$.

\subsection{Varying $\mu$ }

\begin{figure}[htb]
\begin{center}
\includegraphics[scale=0.55]{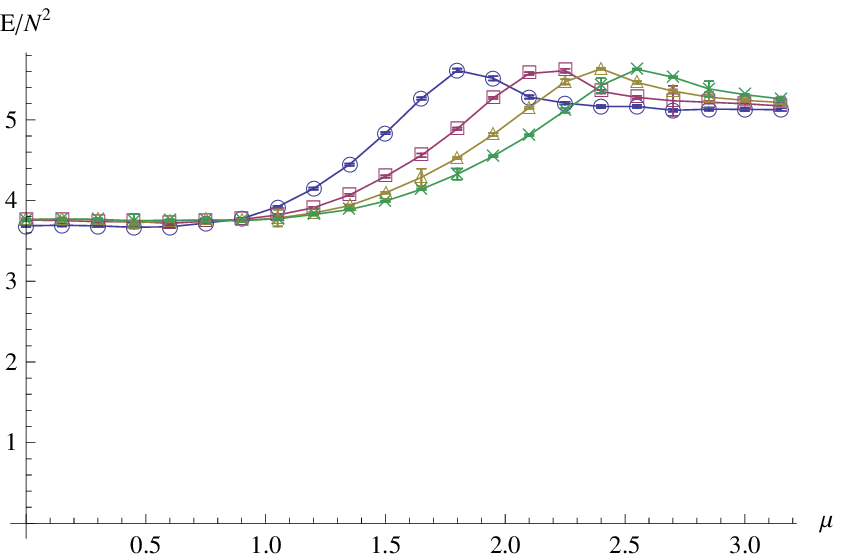}
\includegraphics[scale=0.55]{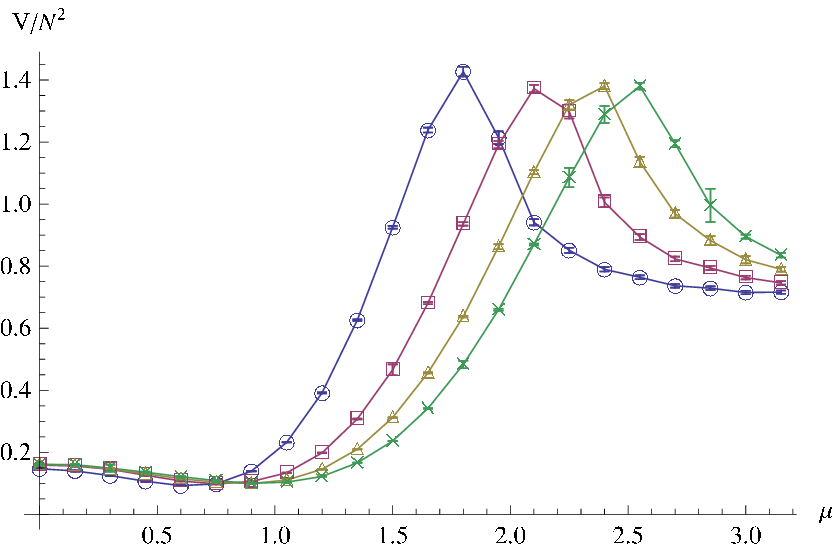}
\includegraphics[scale=0.55]{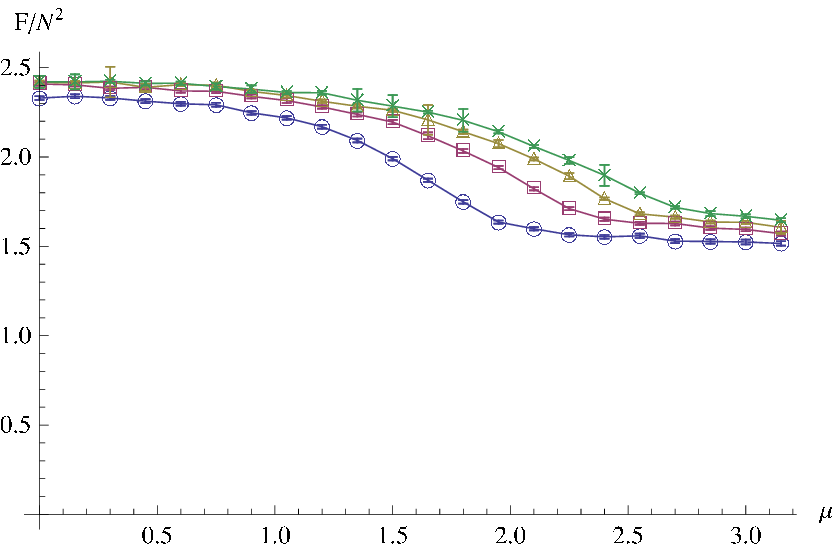}
\includegraphics[scale=0.55]{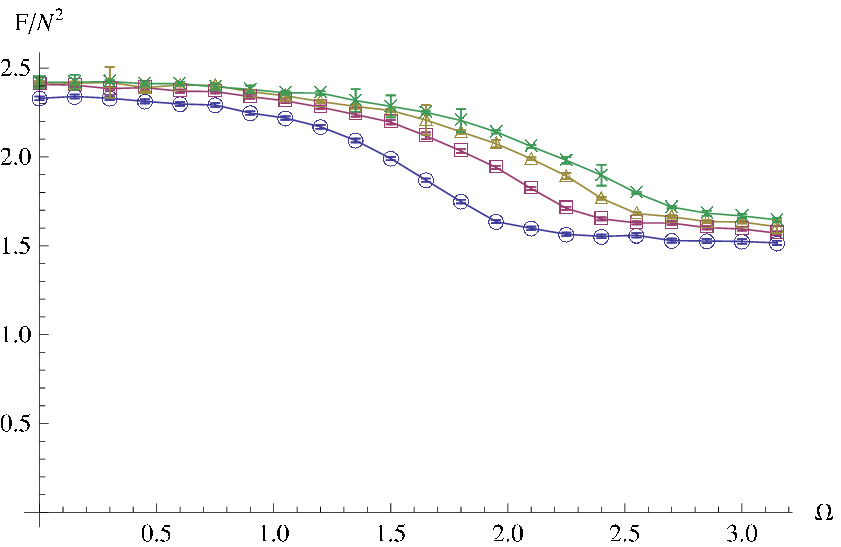}
\includegraphics[scale=0.55]{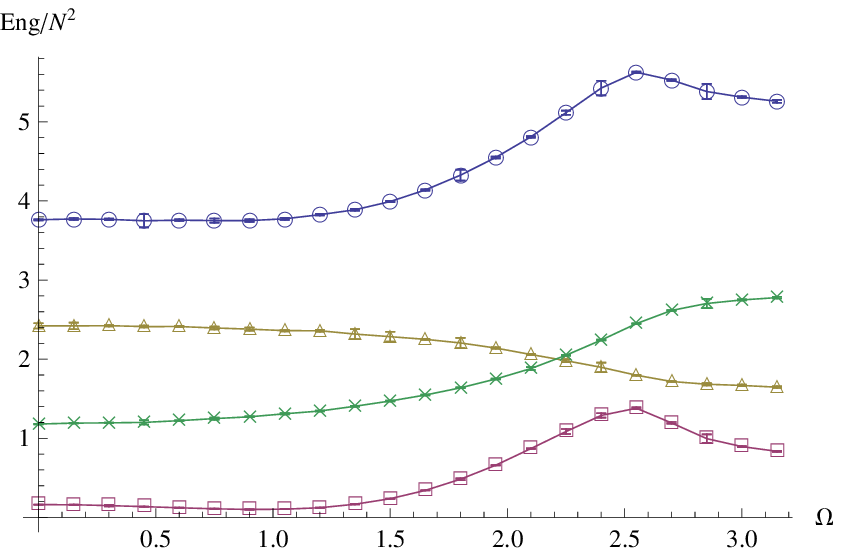}
\end{center}
\caption{\footnotesize The total energy density and the various
  contributions for $\Omega=0$ varying $\mu$ and $N$. From the left to
  the right $E$, $V$, $D$, $F$ an comparison with $N=5$ (circle),
  $N=10$ (square), $N=15$ (triangle), $N=20$ (cross). For the
  comparison: $E$ (circle), $V$ (square), $D$ (triangle), $F$
  (cross).\normalsize}\label{Figure 16}\end{figure}

In this section we analyse the response of the system varying
$\mu\in[0,3]$ while $\Omega$ is fixed at $0$, $1$ or $3$, and $\alpha$ is
always zero. We start displaying the plots fig.\ref{Figure 16} of the
total energy density and of various contributions for $\Omega=0$.
There is no evident discontinuity but there appears a peak in the total
energy density around $\mu\approx 2.5$ for $N=20$. Comparing all the
contributions it is easy to notice that the slope of the total energy is
dictated by the curve $V$ of the potential part.
\begin{figure}[htb]
\begin{center}
\includegraphics[scale=0.7]{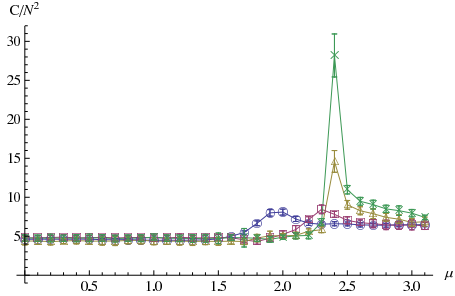}
\end{center}
\caption{\footnotesize Specific heat density for $\Omega=0$ varying
  $\mu$ and $N$.\normalsize}\label{Figure 17}
\end{figure}

\begin{figure}[htb]
\begin{center}
\includegraphics[scale=0.55]{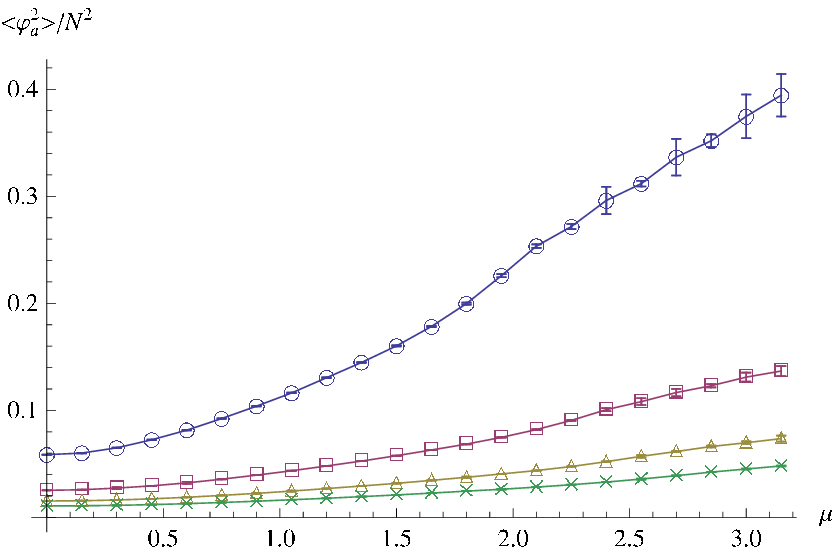}
\includegraphics[scale=0.55]{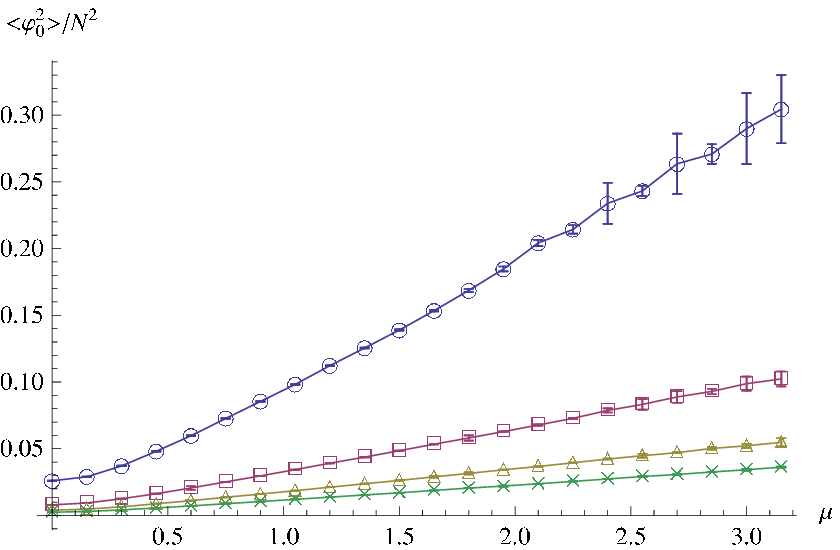}
\includegraphics[scale=0.55]{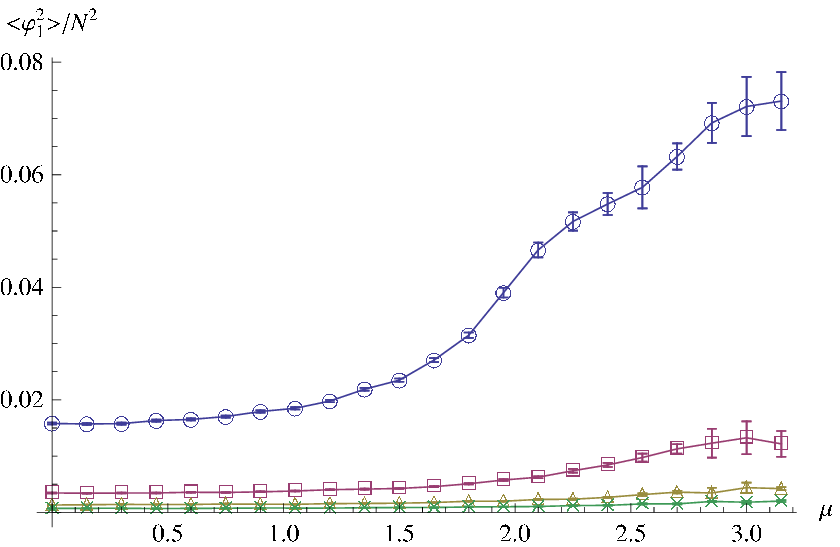}
\includegraphics[scale=0.55]{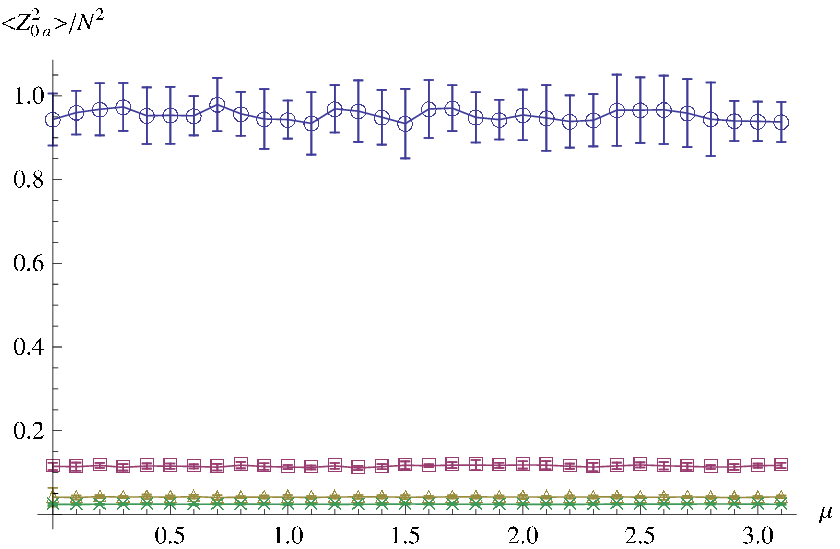}
\includegraphics[scale=0.55]{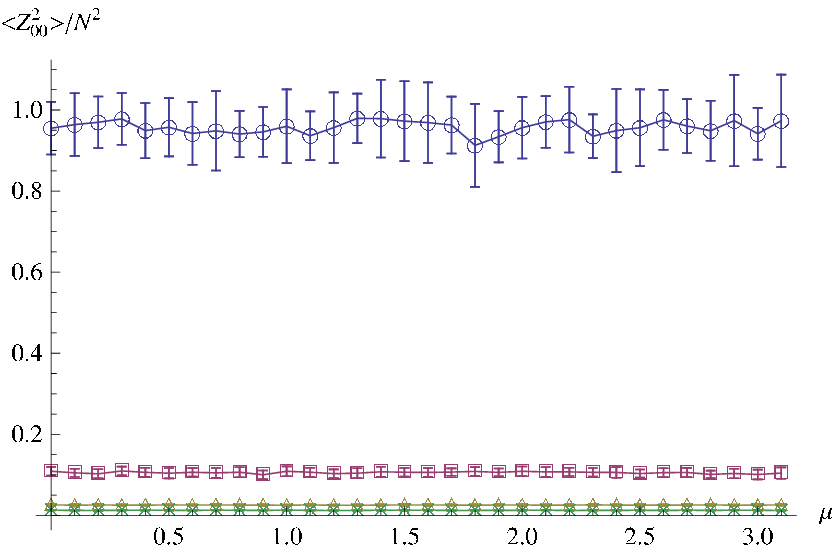}
\includegraphics[scale=0.55]{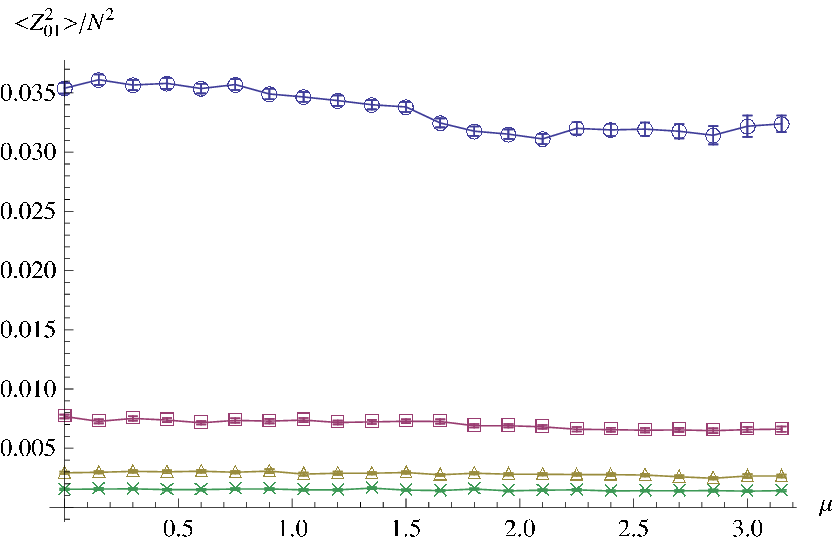}
\end{center}
\caption{\footnotesize Starting from the up left corner and from the
  left to the right the densities for $\langle\varphi_a^2\rangle $,
  $\langle\varphi^2_0 \rangle$, $\langle\varphi^2_1\rangle $, $\langle
  Z_{0a}^2 \rangle$, $\langle Z_{00}^2 \rangle $ and $\langle
  Z_{01}^2\rangle$ for $\Omega=0$ varying $\mu$ and $N$.
  \normalsize}\label{Figure 18}
\end{figure}

As mentioned before, the specific heat density fig.\ref{Figure 17}
features a peak around $\mu\approx2.5$ for $N=20$. Again, since the
peak increases with $N$, we could relate this to a phase transition.
The plots for the quantities $\langle\varphi_a^2\rangle $ and
$\langle\varphi^2_0 \rangle$ show a strong dependence on $\mu$, in
particular the slope of $\langle\varphi^2_0 \rangle$ seems mostly
linear. The plot for $\langle\varphi^2_1\rangle $ also increases with
$\mu$, but not linearly.  From the first three plots of
fig.\ref{Figure 18} we deduce that close to the origin the
non-spherical contribution $\langle\varphi^2_1\rangle $ is bigger than
the spherical one $\langle\varphi^2_0 \rangle$. Increasing $\mu$, this
situation capsizes and $\langle\varphi^2_0 \rangle$ becomes dominant
over $\langle\varphi^2_1\rangle $.

The behaviour of the $Z_0$ fields as shown in the last three plots of
fig.\ref{Figure 18} is quite different.  The spherical contribution
is always dominant for the whole interval $\mu\in[0,3]$. The curves
for $\langle Z_{0a}^2 \rangle$, $\langle Z_{00}^2 \rangle $ are
compatible to the constant slope. For $\langle Z_{01}^2\rangle $ we
have the same dependence on $\mu$, in particular there is a smooth
descending step which becomes smoother for bigger $N$. However, we
admit that due to some cancellation effects, the statistical errors
are quite big so that this interpretation is not fully conclusive.
Anyway, this result demonstrates the dependence of the order parameter
for $Z_i$, and in general of the system, on the two choices $\Omega=0$
or $\Omega\neq0$.

\begin{figure}[htb]
\begin{center}
\includegraphics[scale=0.55]{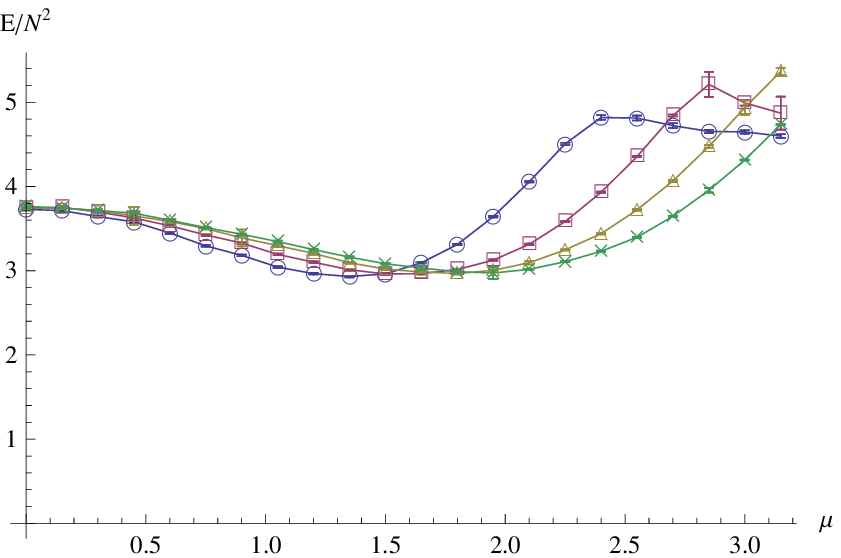}
\includegraphics[scale=0.55]{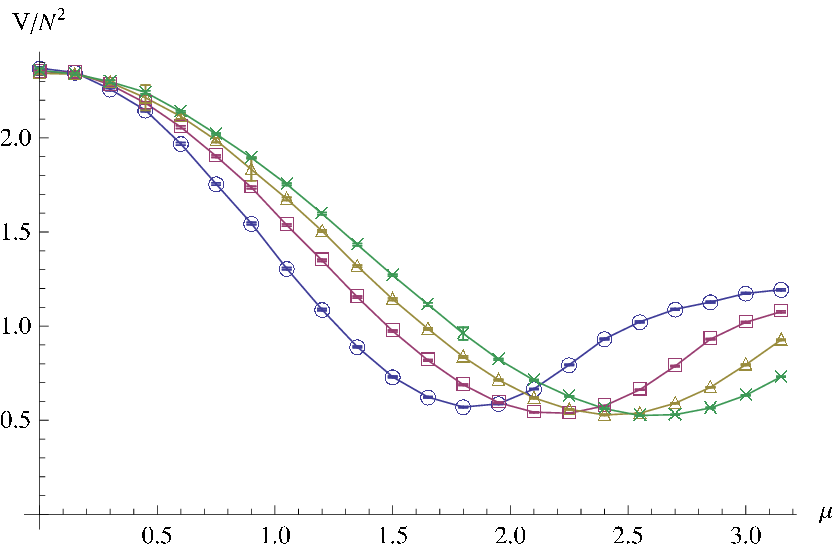}
\includegraphics[scale=0.55]{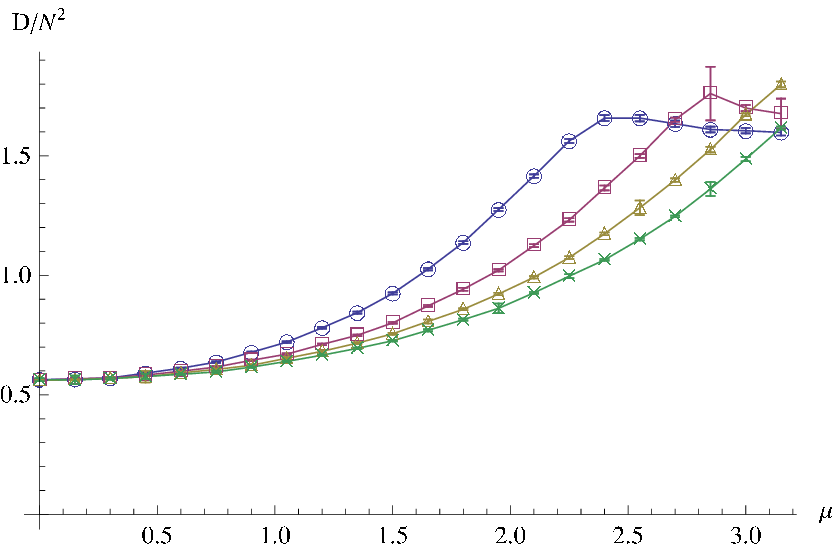}
\includegraphics[scale=0.55]{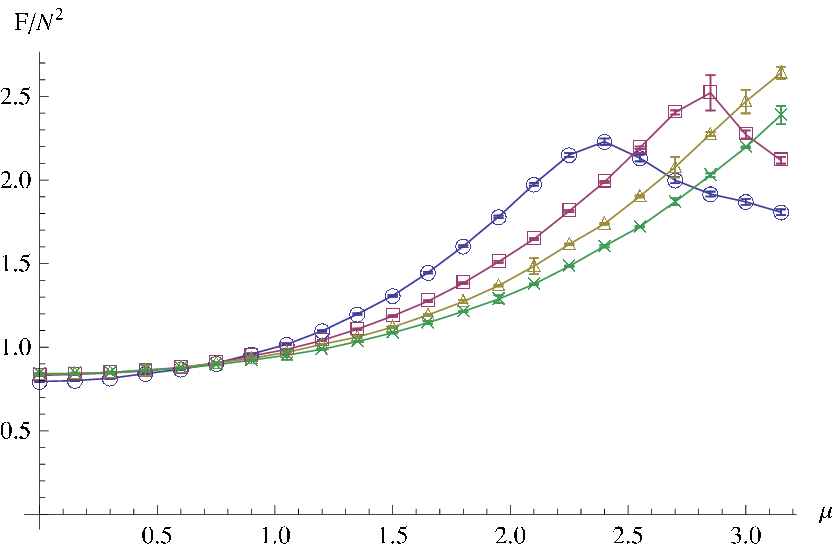}
\end{center}
\caption{\footnotesize Total energy density and contributions for
  $\Omega=1$ varying $\mu$ and $N$. From the left to the right $E$,
  $V$, $D$, $F$.\normalsize}\label{Figure 19}
\end{figure}

\begin{figure}[htb]
\begin{center}
\includegraphics[scale=0.7]{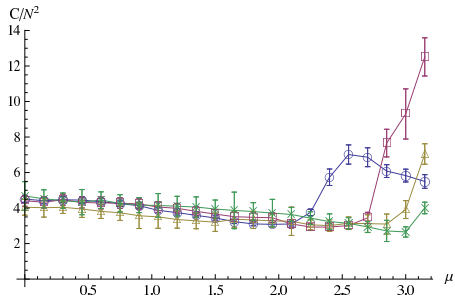}
\end{center}
\caption{\footnotesize Specific heat density for $\Omega=1$ varying
  $\mu$ and $N$.\normalsize}\label{Figure 20}
\end{figure}

Now we will analyse the model for $\Omega=1$. As fig.\ref{Figure 19}
shows, the plots have a different slope compared to the previous case.
The maximum of total energy density follows the one of the
$V$-component. If we focus only on the total energy plot and
compare it with the one for $\Omega=0$, we notice a shift of the
maximum for each $N$. In particular, in fig.\ref{Figure 19} some
maxima are moved outside the considered interval.
We can find this shift very clearly looking at specific heat density
plotted in fig.\ref{Figure 20}. Here again the peak both increases
with $N$ and is shifted to $\mu\approx 3.3$.

\begin{figure}[htb]
\begin{center}
\includegraphics[scale=0.55]{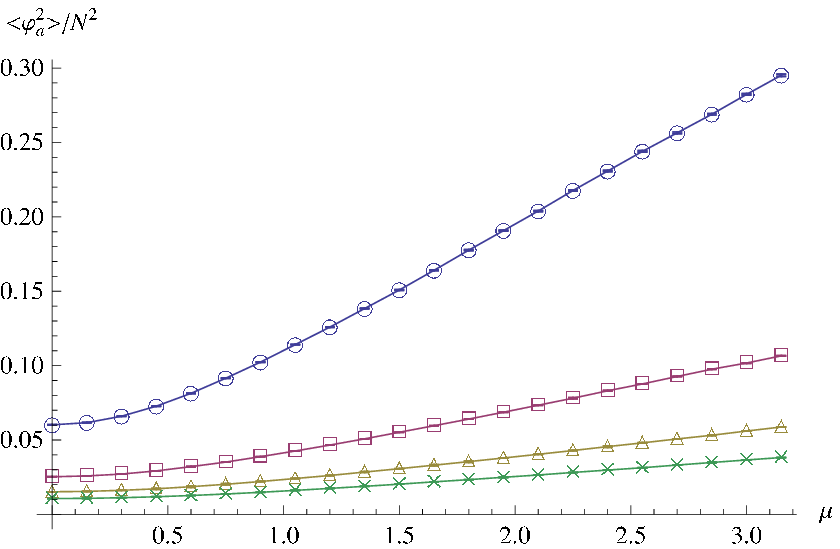}
\includegraphics[scale=0.55]{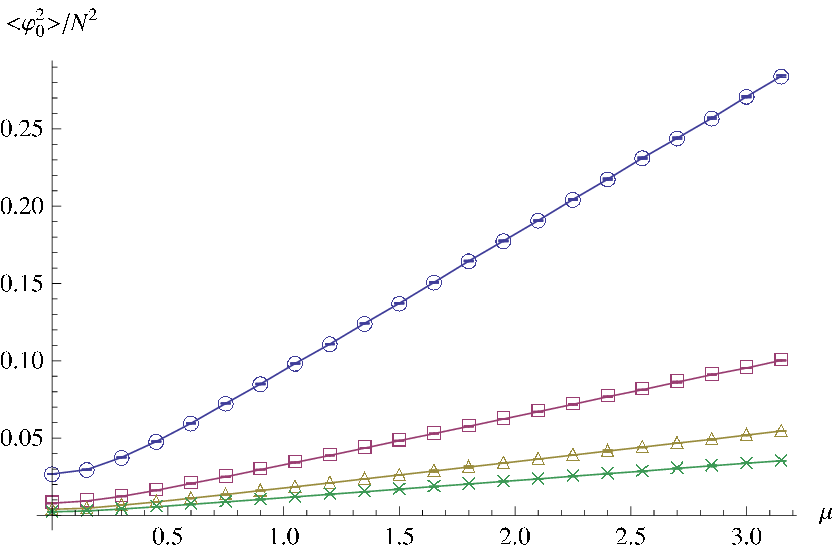}
\includegraphics[scale=0.55]{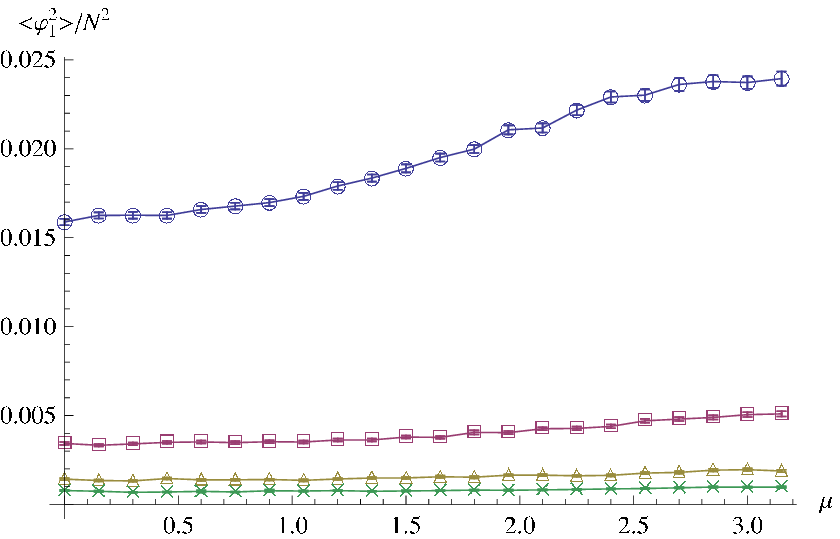}
\includegraphics[scale=0.55]{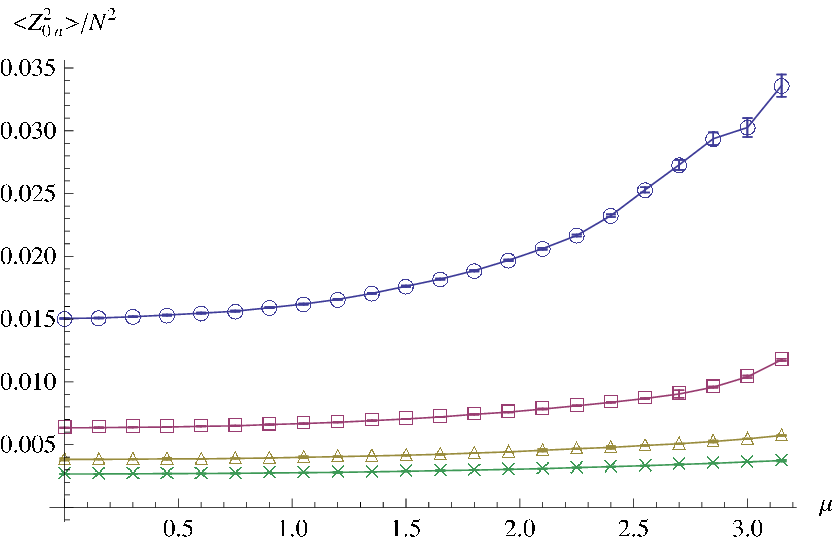}
\includegraphics[scale=0.55]{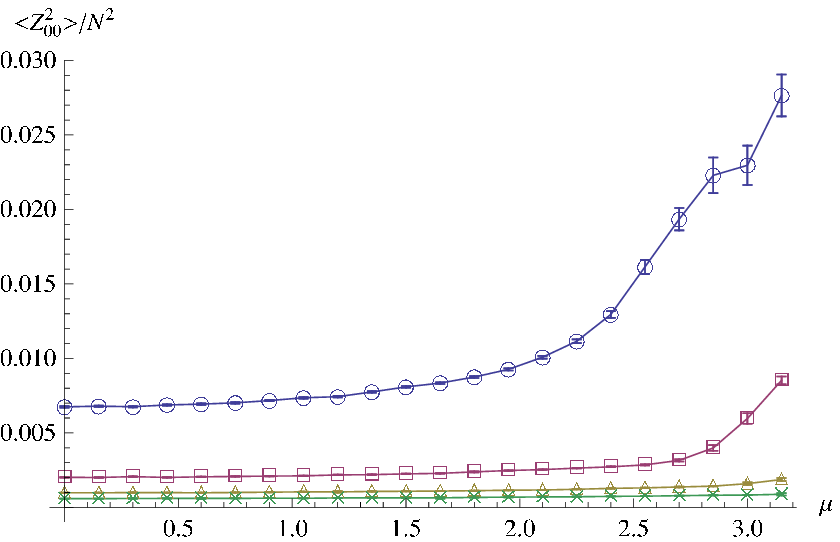}
\includegraphics[scale=0.55]{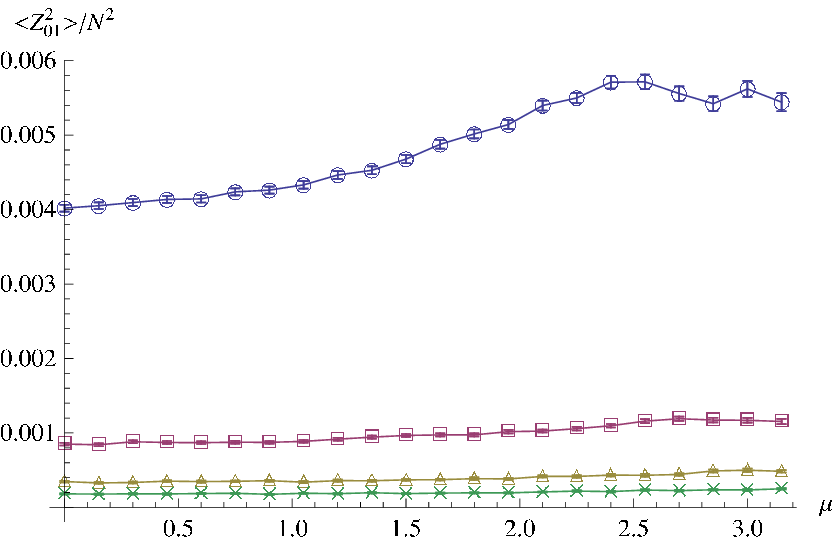}
\end{center}
\caption{\footnotesize Starting from the up left corner and from the
  left to the right the densities for $\langle \varphi_a^2\rangle $,
  $\langle \varphi^2_0 \rangle$, $\langle \varphi^2_1\rangle $,
  $\langle Z_{0a}^2 \rangle$, $\langle Z_{00}^2 \rangle $ and $\langle
  Z_{01}^2\rangle$ for $\Omega=1$ varying $\mu$ and
  $N$.\normalsize}\label{Figure 21}
\end{figure}

Fig.\ref{Figure 21} shows for $\langle \varphi_a^2\rangle $, $\langle
\varphi^2_0 \rangle$ the same behaviour as in the case $\Omega=0$. The
plot for $\langle \varphi^2_1\rangle $ displays an almost constant
curve. However, close to the origin, the spherical contribution and
the first non-spherical one are comparable.
The introduction of $\Omega\neq 0$ creates, in the $Z_0$-order
parameters shown in fig.\ref{Figure 21}, a dependence similar to the
plots for $\psi$. The full-power-of-the-field density and the
spherical contribution are no longer constant, they grow as $\mu$
increases. Even in this case the spherical contribution is always
dominant excluding the region around $\mu=0$.

The last set of plots treats the case $\Omega=3$. The following
diagrams for the energy and its contributions show the absence of the
previous peak. They show a sort of dilatation of the former plots of
fig.\ref{Figure 19}.
\begin{figure}[htb]
\begin{center}
\includegraphics[scale=0.55]{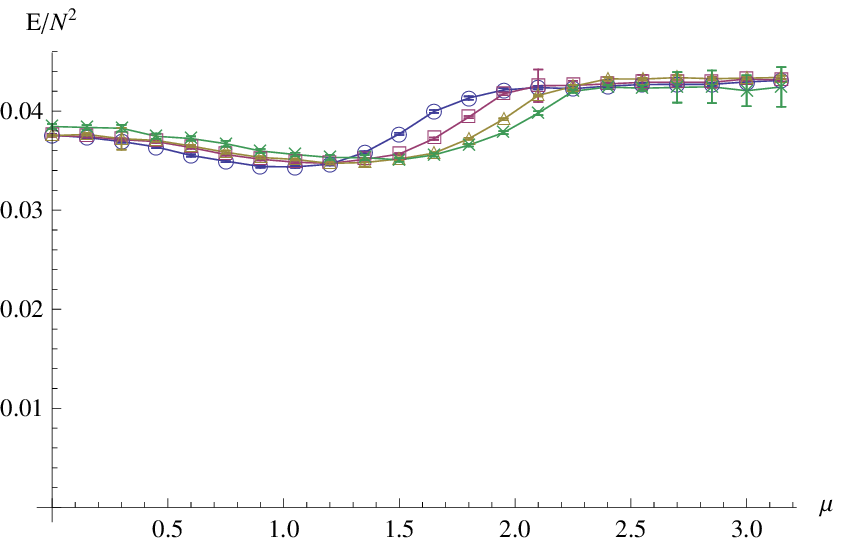}
\includegraphics[scale=0.55]{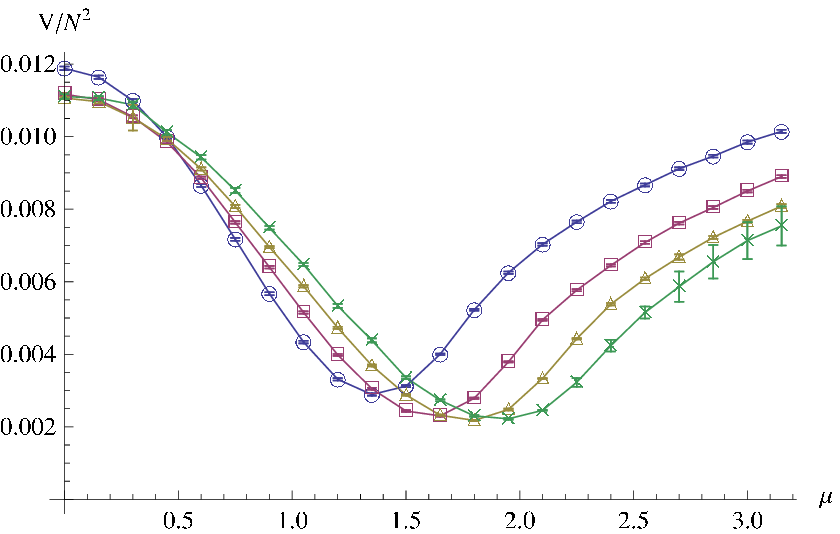}
\includegraphics[scale=0.55]{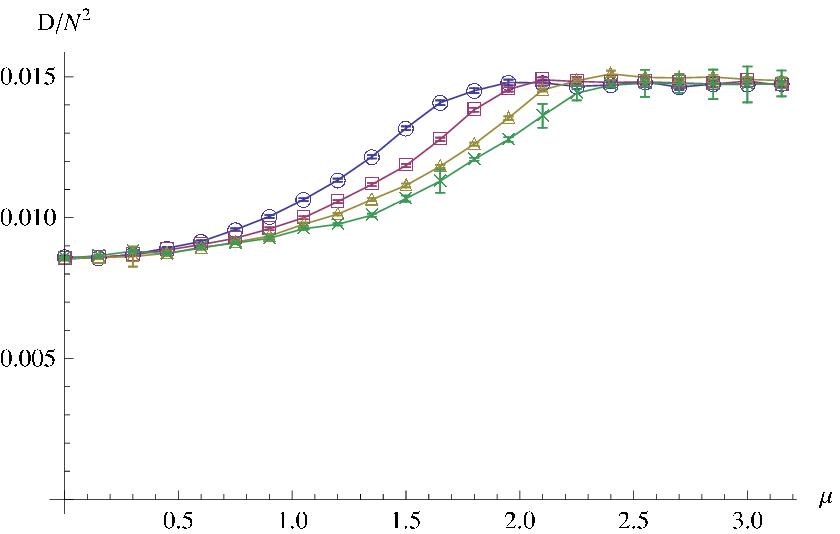}
\includegraphics[scale=0.55]{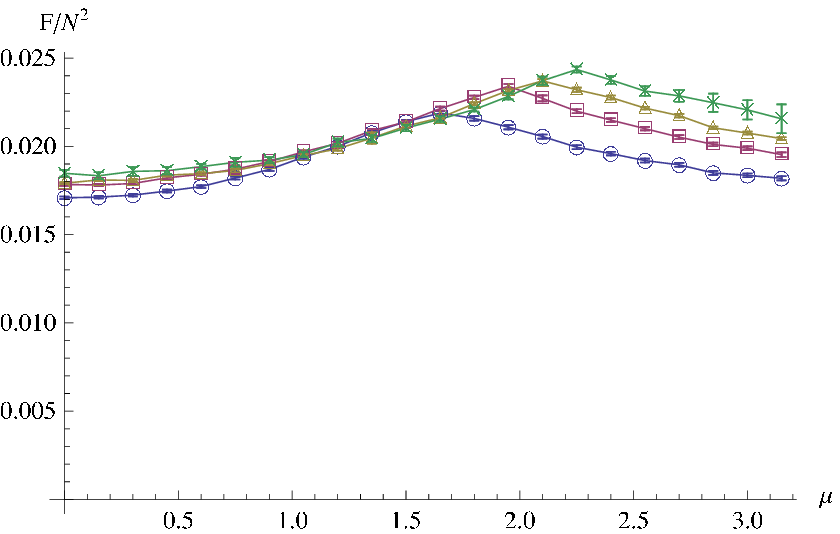}
\end{center}
\caption{\footnotesize Total energy density and the various
  contributions for $\Omega=3$ varying $\mu$ and $N$. From the left to
  the right $E$, $V$, $D$,$F$ . \normalsize}\label{Figure
  22}\end{figure}

\begin{figure}[htb]
\begin{center}
\includegraphics[scale=.8]{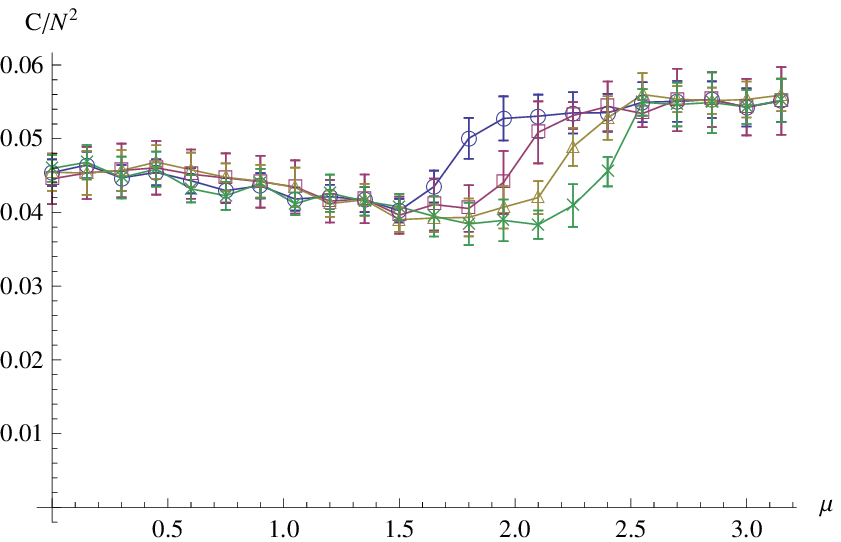}
\end{center}
\caption{\footnotesize Specific heat density for $\Omega=3$ varying
  $\mu$ and $N$.\normalsize}\label{Figure 23}
\end{figure}

The specific heat density fig.\ref{Figure 23} 
does not show the peak in zero anymore,
and the curves do not show any particular point
as $N$ increases. Actually, the peak can be found for higher $\mu$.
\begin{figure}[htb]
\begin{center}
\includegraphics[scale=0.55]{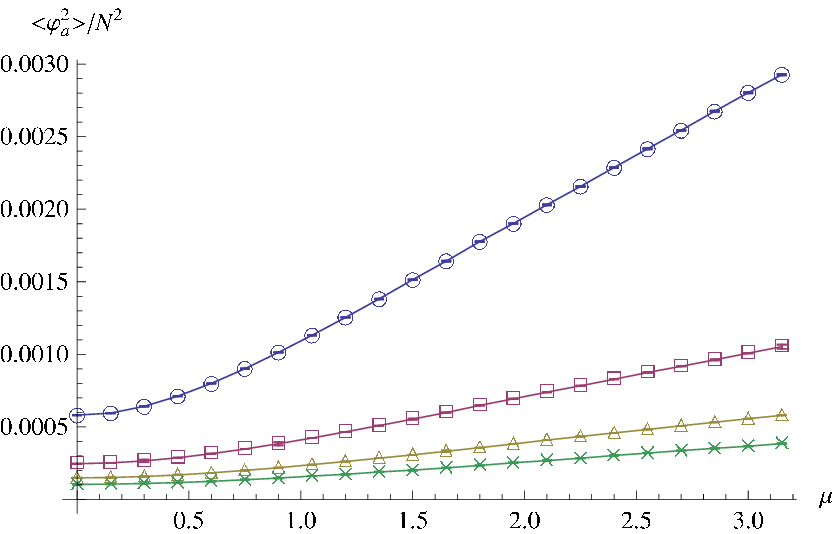}
\includegraphics[scale=0.55]{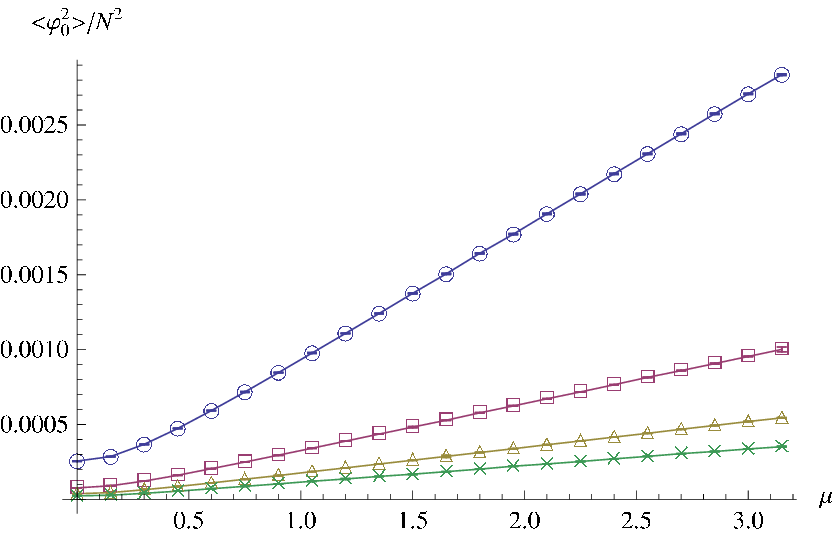}
\includegraphics[scale=0.55]{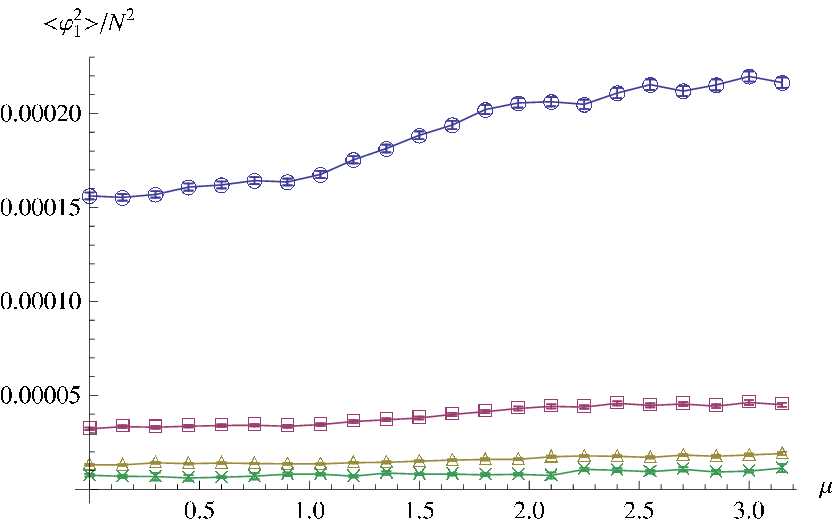}
\includegraphics[scale=0.55]{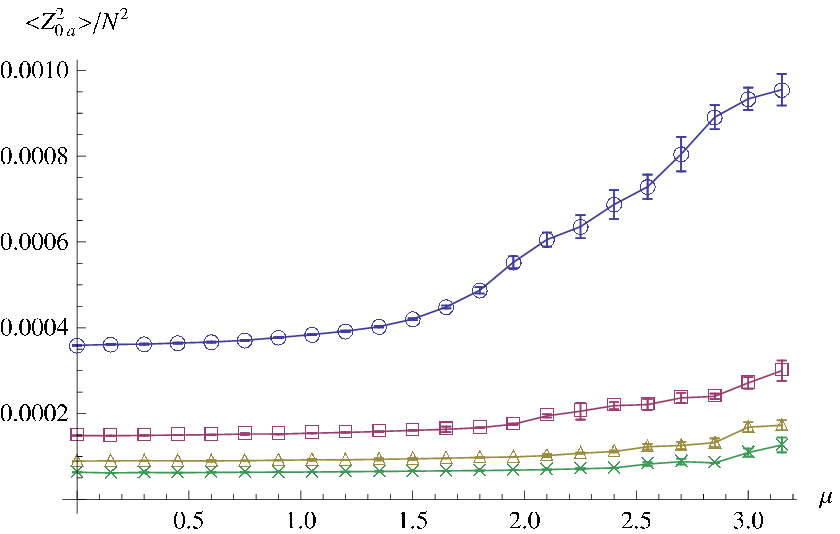}
\includegraphics[scale=0.55]{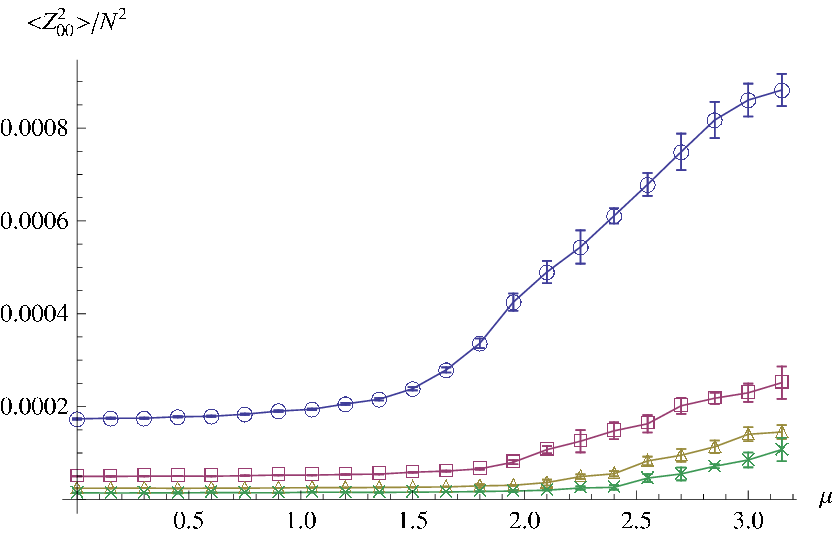}
\includegraphics[scale=0.55]{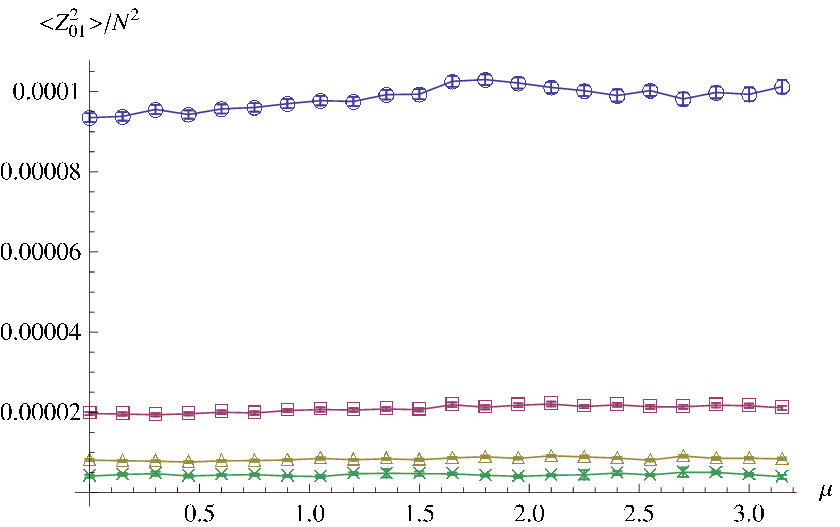}
\end{center}
\caption{\footnotesize Starting from the up left corner and from the
  left to the right the densities for $\langle \varphi_a^2\rangle $,
  $\langle \varphi^2_0 \rangle$, $\langle \varphi^2_1\rangle $,
  $\langle Z_{0a}^2 \rangle$, $\langle Z_{00}^2 \rangle $ and $\langle
  Z_{01}^2\rangle$ for $\Omega=3$ varying $\mu$ and
  $N$.\normalsize}\label{Figure 24}
\end{figure}

At last in fig.\ref{Figure 24} we find a behaviour of the density of
the order parameters for the $Z_0$ and $\psi$ fields similar to the former
plots for $\Omega=1$, and they are compatible with a dilatation of the
previous plots.

\section*{Conclusions and prospectives}

We have studied a noncommutative gauge theory which arises by
restriction of the spectral action for harmonic Moyal space to finite
matrices. For this quantum field theoretical model we have performed
Monte Carlo simulations to obtain, as function of the 
parameters $(\mu,\Omega)$, non-perturbative information for the
energy density, for various contributions to the energy density and
for the specific heat density, as well as for a set of order
parameters related to sphericity.  Despite the complexity of the
approximated spectral action considered here, we were able to obtain
some reliable numerical results, showing that a numerical treatment of
this kind of noncommutative gauge models seems feasible. However, as the
restriction to finite matrices shows severe differences to the
original smooth action, the relevance of our results to the smooth
case is not clear.

The specific heat density shows various peaks which could indicate
phase transitions. In particular, studying the behaviour for some
fixed mass parameter $\mu$ we found a relevant peak close to
$\Omega=0$ for $ \mu=3$, and we noticed a big change in the energy
density and in its contributions between the cases $\mu \in \{0,1\}$
and $\mu=3$. Other peaks in the specific heat density can be found
varying $\mu$ and fixing $\Omega$. The plots show that increasing
$\Omega$, the peak in the specific heat which starts at $\mu\approx
2.4$ for $\Omega=0$ is moved towards higher $\mu$.  The order
parameters we introduced show a strong dependence on the cases
$\Omega=0$ versus $\Omega\neq0$. Referring to the fixed-$\mu$ plots we
found a peak in the spherical contribution for the gauge fields $Z_i$.
Its behaviour can be interpreted as a sort of symmetry breaking
introduced by $\Omega\neq 0$. Additionally, varying $\mu$ and fixing
$\Omega$, the other parameters display a slope increasing with $\mu$
for all fields and all situations but one: the plots of the order
parameters $\langle Z_{0a}\rangle$, $\langle Z_{00}\rangle$ for
$\Omega=0$ show a constant behaviour.

The natural next steps in the numerical study of this model could be
the computation of the transition curves in order to separate the
phase regions and to classify them using possibly additional order
parameters. Our treatment, forced by limited resource, was conducted
varying $\Omega$ in the range $[0,3.1]$, since the Langmann-Szabo
duality does not hold anymore in our case. Actually, the computed
plots do not show any periodicity in $ \Omega\in[0,1]$ so that we can
infer that in contrast to the scalar case the range $[0,1]$ is not
enough to describe the system.  

It will be very interesting to relax the condition $\mu^2 > 0$. Implementing $\mu^2 < 0$ amounts to conduct the calculation no longer around the minimum of the action, in particular  avoiding the explicit use of the finite vacuum  and considering directly the four indexed components we can hope in some improvement in the behavior of the large $N$ limit making the simulations much more harder (but not impossible), however the limit can continue to show big differences. Such continuous limit issues are not new  for the simulations of a scalar field theory on matrix model, in facts in \cite{order-par,order-par1} is discovered a so-called matrix phase, which is not present in the continuous model, as a result the infinite matrix limit fails to converge to the classical case. The adding of an extra term to the matrix action \cite{F-Sphere} solves the problem making the new  	anomalous phase transition to disappear. The extension of the parameters space, together with the classification of the different phase regions, would allow us to compare our model with the results of the simulation performed for the fuzzy sphere, have a look at the occurrence of a correspondent matrix phase and eventually to  try  to regularize the infinite matrix limit.

\section*{Acknowledgements}

This work has been supported by the Marie Curie Research Training
Network MRTN-CT-2006-031962 in Noncommutative Geometry, EU-NCG. Of
particular importance was the interaction with the Dublin Institute
for Advanced Studies.

\end{document}